\documentclass[letterpaper,titlepage,11pt]{article}
\usepackage{amsmath,amssymb,mathtools,bm}
\usepackage{physics}
\usepackage{cancel}
\usepackage{slashed}
\usepackage{mathrsfs}
\usepackage{color}
\usepackage[svgnames]{xcolor}
\usepackage{enumerate}
\usepackage{graphicx,booktabs,array}
\usepackage{tikz}
\usepackage{caption}
\usepackage[symbol]{footmisc}
\usepackage{enumitem}
\usepackage{empheq}

\usepackage{multirow}
\usepackage{booktabs}

\usetikzlibrary{arrows.meta, positioning, calc, decorations.pathmorphing, shapes.geometric}
\usepackage{tikzsymbols}
\usepackage{fontawesome5}
\usepackage{pifont}
\usetikzlibrary{shadows}

\usepackage[numbers,sort&compress]{natbib}
\allowdisplaybreaks

\numberwithin{equation}{section}

\usepackage[
colorlinks=true,
linkcolor=DarkBlue,
anchorcolor=Black,
urlcolor=DarkRed,
filecolor=green,
citecolor=DarkRed,
linktoc=page,
pdfstartview=FitV,
pdftitle={A Ghost-Free dS/AdS Correspondence in N=1 Supergravity},
pdfauthor={Omer Guleryuz},
pdfsubject={},
pdfkeywords={},
bookmarksopen=true
]{hyperref}

\usepackage[nameinlink,capitalise]{cleveref}

\newcommand{\K}{\mathcal{K}}
\newcommand{\W}{\mathcal{W}}
\newcommand{\Lsc}{\mathcal{L}_{\text{sc}}}
\newcommand{\Lf}{\mathcal{L}_{\text{ferm}}}
\newcommand{\eL}{\epsilon_{\mathcal{L}}}
\newcommand{\eg}{\epsilon_g}
\newcommand{\Nstd}{\mathcal{N}_{\text{std}}}
\newcommand{\Xh}{\hat{X}}
\newcommand{\Xhb}{\bar{\hat{X}}}
\newcommand{\RR}{\mathcal{R}}
\newcommand{\dS}{\text{dS}}
\newcommand{\AdS}{\text{AdS}}
\newcommand{\LB}{L_{\mathcal{B}}}

\begin{document}

 \hypersetup{pageanchor=false}
 \begin{titlepage}
			
			\bigskip
			
			\begin{center}
				{\LARGE\bfseries A Ghost-Free dS/AdS Correspondence in $\mathcal{N}=1$ Supergravity}
				\\[10mm]
				\textbf{Omer Guleryuz}\\[5mm]
				\vskip 25pt

				{\em \hskip -.1truecm Department of Physics, Istanbul Technical University, \\
					Maslak 34469 Istanbul, Türkiye \vskip 5pt }

				{{\tt \href{mailto:omerguleryuz@itu.edu.tr}{omerguleryuz@itu.edu.tr}}
 
 }

			\end{center}
			
			\vspace{3ex}

			\begin{center}
				{\bfseries Abstract}
			\end{center}
\begin{quotation}
Reversing the sign of the scalar potential exchanges de Sitter and anti-de Sitter vacua and, through the domain-wall/cosmology correspondence, maps flat domain walls onto flat FLRW cosmologies, but its supersymmetric version has ghosts on the cosmological side. We show that this pointwise correspondence is realized between two ordinary, ghost-free four-dimensional $\mathcal{N}=1$ supergravities with spontaneously broken supersymmetry. Within a single superconformal K\"ahler cone, a $\mathbb{Z}_2$ reflection of the nilpotent goldstino K\"ahler metric through the flat limit $V=0$ reverses $V\leftrightarrow-V$ pointwise while preserving the visible K\"ahler metric, the superpotential, and the unitary-gauge fermion Lagrangian. The paired vacua obey a sharp dictionary with $H=1/L$, $m^2_{\rm dS}=-m^2_{\rm AdS}$, equal conformal weights, and $m_{3/2}>H$, and are simultaneously stable precisely within the window $0\le m^2_{\rm dS}\le\frac94H^2$ set by the Breitenlohner--Freedman bound. Applied to string compactifications and cosmology, this correspondence pairs over-uplifted with under-uplifted anti-D3-brane vacua, confines the de Sitter swampland conjecture to the goldstino (anti-brane) sector, and maps slow-roll inflation onto a healthy AdS domain wall obeying the holographic $c$-theorem.
\end{quotation}
			
			\vfill
			
 \begin{center}{\flushleft{\today}}\end{center}
		\end{titlepage}
		\hypersetup{pageanchor=true}
		\setcounter{page}{1}
		\tableofcontents

\newpage

\section{Introduction}\label{sec:intro}

Whether de Sitter space admits a consistent quantum description, and how it is related to anti-de Sitter space, are central questions of high-energy theory. At the classical level the two are related by a reversal of the scalar potential, $V\to-V$, which exchanges a de Sitter (dS) vacuum with Hubble rate $H$, where $V=3H^2$, and an anti-de Sitter (AdS) vacuum with curvature radius $L$, where $V=-3/L^2$, with $H=1/L$. The domain-wall/cosmology correspondence \cite{Skenderis:2006jq,Skenderis:2006fb,Skenderis:2007sm} promotes this to a map between solutions: every flat domain wall of a theory with potential $V$ is a flat Friedmann--Lema\^itre--Robertson--Walker (FLRW) cosmology of the theory with potential $-V$. The map underlies holographic models of cosmology \cite{McFadden:2009fg}, in which inflationary backgrounds are continuations of holographic renormalization-group flows, and it is closely related to the continuation from AdS/CFT to dS/CFT \cite{Strominger:2001pn,Maldacena:2002vr}.

The asymmetry between the two signs is sharpest in string theory. Flux compactifications readily yield no-scale Minkowski vacua and supersymmetric or non-supersymmetric AdS vacua \cite{Giddings:2001yu,Kachru:2003aw,Balasubramanian:2005zx}, whereas compactifications of classical two-derivative supergravity that satisfy the assumptions of the Maldacena--Nu\~nez theorem, in particular a smooth compact internal space without negative-tension sources, admit no dS vacua \cite{Maldacena:2000mw}. Proposed dS constructions add further ingredients, most prominently an anti-D3-brane that uplifts an AdS vacuum \cite{Kachru:2003aw,Kachru:2002gs}, whose consistency remains under active debate \cite{Danielsson:2018ztv,Obied:2018sgi,Ooguri:2018wrx}. A ghost-free map between dS and AdS vacua allows the two sides of this divide to be confronted directly within a single effective framework.

In supergravity the dS side of the reversal has long been problematic. Unbroken supersymmetry in dS$_4$ cannot be realized on a positive-definite Hilbert space \cite{Pilch:1984aw,Lukierski:1984it}, and in the supersymmetric form of the domain-wall/cosmology correspondence the cosmological partner of a supersymmetric domain wall is a pseudo-supergravity, with twisted reality conditions and vector ghosts \cite{Skenderis:2007sm,Bergshoeff:2007cg}. The question has also been approached from the superconformal formulation of supergravity: in Ref.~\cite{Atli:2020dni} a sign choice in the exponent of the superconformal K\"ahler cone produced a positive potential from a K\"ahler function with negative-definite metric, which suggested that superconformal sign choices might connect AdS and dS potentials. Healthy dS supergravities do exist once supersymmetry is broken spontaneously by a nilpotent goldstino multiplet \cite{Ferrara:2014kva,Bergshoeff:2015tra}, and at constant fields two choices of its coupling already give two such theories with opposite cosmological constants. The domain-wall/cosmology correspondence, however, demands far more: two genuinely healthy supergravities that share the same scalar kinetic geometry while carrying potentials that are exact mirror images, $V \leftrightarrow -V$, across field space. Whether four-dimensional $\mathcal{N}=1$ supergravity can realize this full pointwise reflection without retreating into pseudo-supergravity or introducing ghosts has, to our knowledge, remained an open challenge.

In this paper we show that such ghost-free partner theories not only exist, but emerge naturally from a two-layer mechanism hidden inside a single superconformal K\"ahler cone with two independent controls. The first control consists of the global signs of the superconformal action: its overall sign, the sign of the cone potential, and the sign of the exponent that contains the K\"ahler function. Because the superconformal densities are linear in their arguments, every choice of these signs is equivalent to plus or minus standard supergravity, and the signs only decide which real function plays the role of the K\"ahler potential. They select a ghost-free base geometry, but they cannot reverse its potential: wherever the superpotential $W$ vanishes, a ghost-free theory is locked to $V\ge0$. What unlocks the sign of the potential is the second control: the weight $h$ of a nilpotent goldstino direction of the cone \cite{Rocek:1978nb,Komargodski:2009rz,Ferrara:2014kva,Bergshoeff:2015tra}. At $W\neq0$ the cone has one negative direction, that of the conformal compensator, and positive directions for the visible fields and for the goldstino, so that the scalar potential takes the form
\begin{equation}\label{eq:intro_V}
V_h=\mathcal{B}+\frac{\mathcal{A}}{h},\qquad \mathcal{B}=|F_{\rm vis}|^2-3m_{3/2}^2,\qquad \mathcal{A}=e^{k}|f|^2 .
\end{equation}
Here $k$ is the K\"ahler potential of the visible chiral multiplets, $m_{3/2}$ the gravitino mass, $|F_{\rm vis}|^2$ the F-term energy of the visible fields, $f$ the coupling of the goldstino multiplet in the superpotential and $h$ its K\"ahler metric, all defined precisely in section~\ref{sec:construction}; $\mathcal{A}/h$ is the F-term energy of the goldstino and $\mathcal{B}$ the energy of the visible fields and the compensator. The map
\begin{equation}\label{eq:intro_R}
\RR(h)=\frac{h\,\mathcal{A}}{-2h\,\mathcal{B}-\mathcal{A}}
\end{equation}
is an exact reflection of the goldstino energy about the value $-\mathcal{B}$ at which the potential vanishes. On its domain $\mathcal{B}<-|V_h|$ it reverses the potential pointwise, $V_{\RR(h)}=-V_h$, while keeping the visible K\"ahler metric, the superpotential, the positivity of the goldstino metric and the surviving unitary-gauge fermion operators completely unchanged.

The outcome is a pair of explicit Lagrangians, Eqs.~(\ref{eq:L_dS})--(\ref{eq:L_AdS}), which differ only in the sign of the potential: both are ordinary, ghost-free $\mathcal{N}=1$ supergravities with spontaneously broken supersymmetry. At corresponding vacua their spectra obey a sharp dictionary,
\begin{equation}\label{eq:intro_dict}
H=\frac1L,\qquad m^2_{\dS}=-m^2_{\AdS},\qquad \Delta_{\dS}=\Delta_{\AdS},\qquad m_{3/2}^2\ge H^2=L^{-2},
\end{equation}
where $m^2$ are the squared masses of corresponding scalar modes and $\Delta$ the conformal weights that govern their late-time falloff in dS and their near-boundary falloff in AdS. The last relation is the Deser--Waldron unitarity bound for the massive gravitino of the AdS member \cite{Deser:2001pe}; through the common gravitino mass it is transferred to the dS member, where it takes the form of the bound $H\lesssim m_{3/2}$ known from KKLT \cite{Kallosh:2004yh}. The second relation turns the Breitenlohner--Freedman (BF) bound \cite{Breitenlohner:1982jf} of the AdS member into the complementary-series bound of the dS member, so that both vacua are linearly stable precisely when
\begin{equation}\label{eq:intro_window}
0\le m^2_{\dS}\le\frac94H^2
\end{equation}
for every scalar mode. Critical points are paired with reversed curvature: a dS minimum in this window is paired with a BF-stable AdS maximum, an inflationary dS hilltop with a genuine AdS minimum, and a Minkowski minimum with a Minkowski maximum. Every flat domain wall of the AdS member is a flat FLRW cosmology of the dS member. To our knowledge, this provides the first realization of a pointwise $V\leftrightarrow-V$ dS/AdS correspondence--and of the domain-wall/cosmology correspondence--in which both members are ordinary, ghost-free $\mathcal{N}=1$ supergravities sharing the same matter and unitary-gauge fermionic sectors.

This dictionary has immediate consequences across cosmology and string compactifications. The weights $\Delta$ that coincide on the two sides are the exponents of the squeezed limit of inflationary correlators: for a tachyon-free dS vacuum a stable AdS partner exists exactly when they are real, and the oscillatory squeezed-limit signals that heavy fields can produce \cite{Chen:2009zp,Arkani-Hamed:2015bza} are the dS image of a BF violation. In flux compactifications $\mathcal{B}$ is the energy of the flux background and $\mathcal{A}/h$ the energy of an anti-D3-brane, so that $\RR$ pairs over-uplifted dS vacua with under-uplifted AdS vacua. The stability window then separates two classes of string-derived dS vacua. Vacua with moduli heavier than $\frac32H$ have ghost-free but BF-unstable partners, while a stable partner requires every scalar to satisfy
\begin{equation}\label{eq:intro_crit}
m^2<\frac{9}{4\LB^2}\le\frac94\,m_{3/2}^2,
\end{equation}
where the length $\LB$ is defined by $\mathcal{B}=-3/\LB^2$. This is a condition on the background functions at the vacuum, into which the uplift enters only through the position of the vacuum. In one-modulus KKLT, closed-form expressions show that in the controlled regime, where the instanton action $a\sigma$ is large, both moduli exceed the bound (\ref{eq:intro_crit}) by a factor of order $(a\sigma)^2$ for every uplift for which a partner exists. In inflationary cosmology the window applies at the final vacuum, where it constrains the curvature of the potential but not the slope that ends inflation. We construct a closed-form inflationary pair with a slow-roll plateau, $n_s\simeq0.97$ and a graceful exit into a dS minimum whose partner is a BF-stable AdS maximum, while the cosmological trajectory between them maps to a healthy AdS domain wall that obeys the classical holographic $c$-theorem. The window then implies that the asymptotic linear relaxation of the inflaton in the quadratic basin of this minimum is overdamped, whereas final vacua with harmonic oscillations, $m>\frac32H$, have BF-unstable partners; transient anharmonic oscillations outside the basin are not constrained. Moreover, because the two members share $(k,w,f)$ and $|V|$ with all its normalized derivatives, the de Sitter swampland conjecture is isolated within this class to the goldstino (anti-brane) sector alone.

It is instructive to contrast this correspondence with two familiar ingredients in the literature. First, solving for the nilpotent metric as $h=\mathcal{A}/(V-\mathcal{B})$ realizes a prescribed potential within a single theory; this is Eq.~(2.12) of Ref.~\cite{Kallosh:2017wnt} (see also Ref.~\cite{McDonough:2016der}), where it is used to build inflationary models one theory at a time. Second, the dS counterpart $m^2\le\frac94H^2$ of the BF bound was noted in the context of the domain-wall/cosmology correspondence in Refs.~\cite{Skenderis:2006rr,Skenderis:2006fb}, where it separates dS vacua approached monotonically from those approached with oscillations. The reflection (\ref{eq:intro_R}) introduced here is of a fundamentally different nature: it relates two theories rather than reconstructing one. By conjugating the reflection $V\mapsto-V$ of potential values with single-theory reconstruction, it acts as a genuine $\mathbb{Z}_2$ reflection on the space of nilpotent K\"ahler metrics through the flat theory $h_*=-\mathcal{A}/\mathcal{B}$ with identically vanishing potential, linking two healthy supergravities with the same $(k,w,f)$--and hence the same visible K\"ahler metric and surviving fermion operators (section~\ref{sec:lagrangians})--but opposite field-dependent potentials. All our results are exact statements at the level of classical effective actions: superconformal density identities, potentials after elimination of the auxiliary fields, fermion operators in unitary gauge, and symmetry-reduced actions. Establishing a non-perturbative or holographic equivalence between the two members would require further ultraviolet and boundary data.

The paper is organized as follows. Section~\ref{sec:construction} constructs the correspondence: the superconformal cone, its two controls and the healthy dS and AdS Lagrangians. Section~\ref{sec:dictionary} develops the dS/AdS dictionary, section~\ref{sec:holography} relates the construction to holographic cosmology, dS/CFT and cosmological correlators, and section~\ref{sec:example} gives an explicit pair. Section~\ref{sec:strings} applies the construction to flux compactifications with an anti-D3-brane uplift and to swampland criteria, and section~\ref{sec:inflation} to inflationary cosmologies, including a closed-form inflationary pair that realizes all three pairings of critical points. Section~\ref{sec:conclusions} contains our conclusions. Appendix~\ref{app:W0} shows that on loci with $W=0$ no covariant rearrangement of the K\"ahler potential and the superpotential reverses the potential of a ghost-free theory, and appendix~\ref{app:stabilizers} collects the spectra of stabilizer models and a profile with separate Minkowski and de Sitter minima.\footnote{We use signature $(-,+,+,+)$, units in which the reduced Planck mass is $M_{\text{Pl}}=1$, and curvature conventions in which the Ricci scalar $R$ of de Sitter space is $R=+12H^2$; $e\equiv\sqrt{-\det g_{\mu\nu}}$ is the volume density. Curvature-sign conventions vary across the supergravity literature (see e.g.\ \cite{Kallosh:2000ve}), and comparing component formulas across conventions is a further source of apparent sign paradoxes.}

\section{The Superconformal dS/AdS Correspondence}\label{sec:construction}

This section constructs the pairs in three steps. A superconformal cone with a nilpotent goldstino direction contains two independent controls (section~\ref{sec:cone}). Its signs select a healthy base geometry but leave the potential locked on $W=0$ loci (section~\ref{sec:signs}). The weight of the goldstino direction balances the compensator at $W\neq0$, and reflecting it reverses the potential (section~\ref{sec:reflection}). The result is a pair of explicit healthy Lagrangians with opposite potentials (section~\ref{sec:lagrangians}).

\subsection{The superconformal cone with a goldstino direction}\label{sec:cone}

In the superconformal approach \cite{Kaku:1978nz,Cremmer:1982en,Kugo:1982cu,Ferrara:2010in,Freedman:2012zz} the coupling of $n$ chiral multiplets to $\mathcal{N}=1$ supergravity is described by $n+1$ chiral multiplets $X^I=(X^0,X^a)$ of Weyl weight one, with $I=0,\dots,n$ and $a=1,\dots,n$, where $X^0$ is the conformal compensator. The theory is specified by a real K\"ahler cone potential $\mathcal{N}(X,\bar X)$, homogeneous of degree one in $X$ and in $\bar X$ separately, and a holomorphic superpotential $\W(X)$, homogeneous of degree three. The action is the sum of a D-term and an F-term density of the superconformal tensor calculus,
\begin{equation}\label{eq:Lsc}
\Lsc[\mathcal{N},\W]=\big[\mathcal{N}\big]_D+\Big(\big[\W\big]_F+\text{h.c.}\Big),
\end{equation}
where h.c.\ denotes the hermitian conjugate, and both densities are linear in their arguments. The physical scalars are the projective coordinates $z^a$, defined by
\begin{equation}\label{eq:proj}
X^a=\frac{X^0z^a}{\sqrt3}.
\end{equation}
We add a nilpotent chiral multiplet $X^{\rm nil}$ of Weyl weight one, $(X^{\rm nil})^2=0$ \cite{Rocek:1978nb,Komargodski:2009rz,Ferrara:2014kva,Bergshoeff:2015tra}, and define the projective goldstino superfield
\begin{equation}\label{eq:Xhat}
\Xh=\frac{\sqrt3\,X^{\rm nil}}{X^0},\qquad \Xh^2=0 .
\end{equation}
Its lowest component is a bilinear of the goldstino fermion and does not propagate \cite{Komargodski:2009rz}.

We consider the family of cone potentials and superpotentials
\begin{equation}\label{eq:Nsigma}
\mathcal{N}_{\sigma,h}=\sigma_1|X^0|^2\exp\!\Big(\frac{\sigma_2}{3}\K(z,\bar z)\Big)\Big(1-\frac{h(z,\bar z)}{3}\,\Xh\Xhb\Big),\qquad
\W=\tau\Big(\frac{X^0}{\sqrt3}\Big)^3\big(w(z)+\Xh f(z)\big),
\end{equation}
with signs $\sigma_1,\sigma_2=\pm1$, a real function $\K(z,\bar z)$ on the visible field space, which we call the K\"ahler function, a real function $h(z,\bar z)$, holomorphic functions $w(z)$ and $f(z)$, and a real non-zero constant $\tau$. An overall sign $\eL=\pm1$ multiplies the action. Since $z^a$ and $\Xh$ are invariant under dilatations and under the $U(1)$ R-symmetry of the superconformal algebra, $\mathcal{N}_{\sigma,h}$ and $\W$ have the required homogeneity. Because $(\Xh\Xhb)^2=0$, the last factor of $\mathcal{N}_{\sigma,h}$ is the most general dependence on $\Xh\Xhb$; we work in this block-diagonal class, without terms linear in $\Xh$ in the cone potential and without vector multiplets.

The family contains two kinds of data. The signs $(\eL,\sigma_1,\sigma_2)$ act on the cone as a whole, while the weight $h$ acts only on its goldstino direction. The first kind fixes the ghost-free base geometry, and the second controls the sign of the potential.

\subsection{First control: signs and the healthy base}\label{sec:signs}

Standard Poincar\'e supergravity with K\"ahler potential $K$ corresponds to the cone potential
\begin{equation}\label{eq:Nstd}
\Nstd[K]=-|X^0|^2e^{-K/3}.
\end{equation}
Since $\Xh$ is nilpotent, $e^{-h\Xh\Xhb/3}=1-h\Xh\Xhb/3$, so that Eq.~(\ref{eq:Nstd}) gives the first and the linearity of the densities in Eq.~(\ref{eq:Lsc}) the second of the two identities
\begin{equation}\label{eq:identities}
\mathcal{N}_{\sigma,h}=-\sigma_1\Nstd\big[{-\sigma_2}\K+h\Xh\Xhb\big],\qquad \eL\Lsc[\mathcal{N},\W]=\Lsc[\eL\mathcal{N},\eL\W].
\end{equation}
For the complete action, including fermions and auxiliary fields and before any gauge fixing, they give
\begin{equation}\label{eq:dictionary}
\begin{aligned}
\eL\,\Lsc\big[\mathcal{N}_{\sigma,h},\W\big]&=\Lsc\big[\eg\Nstd[K],\eL\W\big]=\eg\,\Lsc\big[\Nstd[K],\eg\eL\W\big]\\
&=\eg\,\Lsc\Big[\Nstd[K],\Big(\frac{X^0}{\sqrt3}\Big)^3W\Big],
\end{aligned}
\end{equation}
with
\begin{equation}\label{eq:dictionary2}
\eg=-\eL\sigma_1,\qquad K=k+h\,\Xh\Xhb,\qquad k=-\sigma_2\K,\qquad W=-\sigma_1\tau\big(w+\Xh f\big).
\end{equation}
The first step uses both identities, $\eL\mathcal{N}_{\sigma,h}=-\eL\sigma_1\Nstd[K]=\eg\Nstd[K]$, and the second the linearity once more, with $\eg^2=1$. In the last step $\eg\eL=-\sigma_1$, so that $\eg\eL\W=(X^0/\sqrt3)^3\,W$ with the superpotential $W$ of Eq.~(\ref{eq:dictionary2}). The right-hand side is $\eg$ times standard supergravity with K\"ahler potential $K$ and superpotential $W$. The constant $\tau$ only rescales $w$ and $f$, and we set $\tau=-\sigma_1$ from now on, so that $W=w+\Xh f$.

The superconformal gauge is fixed by the Einstein gauge $\Nstd[K]=-3$ for the dilatations and by the standard conditions for the special conformal transformations, the S-supersymmetry and the $U(1)$ R-symmetry \cite{Freedman:2012zz}. Along the bosonic locus $\Xh=0$ the bosonic Lagrangian of the right-hand side of Eq.~(\ref{eq:dictionary}) is then
\begin{equation}\label{eq:Einstein}
e^{-1}\mathcal{L}=\eg\Big[\frac12R-k_{a\bar b}\,\partial_\mu z^a\partial^\mu\bar z^{\bar b}-V\Big],\qquad
V=e^K\big(K^{i\bar j}D_iW\,\overline{D_jW}-3|W|^2\big)\Big|_{\Xh=0}.
\end{equation}
Here $k_{a\bar b}\equiv\partial_a\partial_{\bar b}k$ is the K\"ahler metric of the visible fields, the indices $i,j$ run over all chiral multiplets $(z^a,\Xh)$, $K^{i\bar j}$ is the inverse of $K_{i\bar j}\equiv\partial_i\partial_{\bar j}K$, and $D_iW\equiv\partial_iW+(\partial_iK)W$ is the K\"ahler-covariant derivative. The overall sign $\eg$ does not affect the classical field equations: at a constant scalar configuration the Ricci tensor obeys $R_{\mu\nu}=\Lambda g_{\mu\nu}$ with $\Lambda=V$ for either sign. A quantum theory free of ghosts requires a positive overall sign and positive-definite kinetic metrics,
\begin{equation}\label{eq:healthy}
\eg=-\eL\sigma_1=+1,\qquad k_{a\bar b}=-\sigma_2\K_{a\bar b}>0,\qquad h>0,
\end{equation}
where $\K_{a\bar b}\equiv\partial_a\partial_{\bar b}\K$. For a K\"ahler function with negative-definite metric, $\K_{a\bar b}<0$, the healthy branch is $\sigma_2=+1$ with $\eL=-\sigma_1$, that is $(\eL,\sigma_1,\sigma_2)=(-,+,+)$ or $(+,-,+)$, on which $k=-\K$. This is how the choice $\sigma_2=+1$ in Ref.~\cite{Atli:2020dni} produced a positive potential from such a K\"ahler function: on the healthy branch the theory is standard supergravity with $K=-\K$. Because there is a single compensator, the signs act on the whole visible sector through the one global sign $-\sigma_2$. By Sylvester's law of inertia, a K\"ahler function whose metric has mixed signature therefore cannot be made ghost-free by any choice of $(\eL,\sigma_1,\sigma_2)$.

The signs thus fix the healthy base geometry of a given K\"ahler function, but they cannot reverse its potential. Wherever $W=0$ the compensator term $-3e^K|W|^2$ is absent, and in any theory with positive-definite K\"ahler metric and gauge kinetic function
\begin{equation}\label{eq:W0pos}
V\big|_{W=0}=e^KK^{i\bar j}\partial_iW\,\overline{\partial_jW}+V_D\ \ge\ 0,
\end{equation}
where $V_D\ge0$ is the D-term potential of vector multiplets, if present. On $W=0$ loci the sign of the potential is therefore locked to the sign of a kinetic term; the upper block of table~\ref{tab:signs} illustrates this lock, and appendix~\ref{app:W0} shows that no covariant rearrangement of $(K,W)$ that keeps $W=0$ evades it. The negative-potential member of a ghost-free pair therefore requires $W\neq0$. In the family constructed below the two members share the superpotential $w$, so that the reversal acts at $W\neq0$ on both branches.

\subsection{Second control: the goldstino weight}\label{sec:reflection}

On the healthy branch $\eL\mathcal{N}_{\sigma,h}=\Nstd[K]$ with $K=k+h\Xh\Xhb$, and the Poincar\'e formula in Eq.~(\ref{eq:Einstein}) already gives the potential. At $\Xh=0$ the K\"ahler metric is block diagonal, $K_{i\bar j}=\text{diag}(k_{a\bar b},h)$, so that $K^{\Xh\bar{\hat X}}=1/h$ and $K^{a\bar{\hat X}}=0$, while $D_aW=D_aw$ and $D_{\Xh}W=f$, with $D_aw\equiv\partial_aw+k_aw$ and $k_a\equiv\partial_ak$. Equation~(\ref{eq:Einstein}) then gives directly the potential (\ref{eq:VAB}) below, with $k^{a\bar b}$ the inverse of $k_{a\bar b}$. The superconformal cone shows the origin of the three terms before gauge fixing: each direction of the cone contributes to the potential with the sign of its metric. With $q\equiv e^{-k/3}$ the healthy cone potential is $\eL\mathcal{N}_{\sigma,h}=-|X^0|^2q\,(1-h\Xh\Xhb/3)$. In the holomorphic coordinates $(X^0,z^a,\Xh)$ and at the bosonic locus $\Xh=0$, the non-vanishing components of $G_{I\bar J}\equiv\partial_I\partial_{\bar J}\big(\eL\mathcal{N}_{\sigma,h}\big)$ are
\begin{equation*}
G_{0\bar0}=-q,\quad G_{0\bar b}=\frac13\,q\bar X^0k_{\bar b},\quad G_{a\bar b}=|X^0|^2q\Big(\frac13k_{a\bar b}-\frac19k_ak_{\bar b}\Big),\quad G_{\Xh\bar{\hat X}}=\frac13|X^0|^2q\,h,
\end{equation*}
together with $G_{b\bar0}=\overline{G_{0\bar b}}$, and completing the square in $dX^0$ gives the cone metric
\begin{equation}\label{eq:nilcone}
\partial_I\partial_{\bar J}\big(\eL\mathcal{N}_{\sigma,h}\big)\,dX^Id\bar X^{\bar J}=-q\,\Big|dX^0-\frac{X^0}{3}\,k_a\,dz^a\Big|^2+\frac{|X^0|^2q}{3}\Big(k_{a\bar b}\,dz^ad\bar z^{\bar b}+h\,d\Xh\,d\Xhb\Big),
\end{equation}
where the indices $I,J$ now include the goldstino direction. The metric has one negative direction, that of the compensator, and positive directions for the visible fields and the goldstino. In the basis of one-forms $e^0\equiv dX^0-\frac{X^0}{3}k_a\,dz^a$, $dz^a$ and $d\Xh$ it is block diagonal, with inverse $\text{diag}\big(-q^{-1},\,3k^{a\bar b}/(|X^0|^2q),\,3/(|X^0|^2q\,h)\big)$. Since $dX^0=e^0+\frac{X^0}{3}k_a\,dz^a$, the differential of the superpotential at $\Xh=0$ is
\begin{equation*}
d\W=\tau\Big[\frac{(X^0)^2}{\sqrt3}\,w\,e^0+\frac{(X^0)^3}{3\sqrt3}\big(D_aw\,dz^a+f\,d\Xh\big)\Big],\qquad\tau^2=1.
\end{equation*}
Eliminating the auxiliary fields contracts the derivatives $\W_I\equiv\partial\W/\partial X^I$ of the superpotential with the inverse cone metric \cite{Ferrara:2010in}. In this basis the contraction is a sum of three terms, each with the sign of its metric,
\begin{equation}\label{eq:conepot}
V=\big(\eL\mathcal{N}_{\sigma,h}\big)^{I\bar J}\,\W_I\overline{\W_J}=\frac{|X^0|^4}{9q}\Big[k^{a\bar b}D_aw\,\overline{D_bw}+\frac{|f|^2}{h}-3|w|^2\Big],
\end{equation}
where $\big(\eL\mathcal{N}_{\sigma,h}\big)^{I\bar J}$ is the inverse of $G_{I\bar J}$. The Einstein gauge $|X^0|^2q=3$ turns the prefactor into $e^k$, and the potential becomes \cite{Ferrara:2014kva,McDonough:2016der}
\begin{equation}\label{eq:VAB}
V_h=\mathcal{B}+\frac{\mathcal{A}}{h},\qquad \mathcal{B}\equiv|F_{\rm vis}|^2-3m_{3/2}^2,\qquad \mathcal{A}\equiv e^k|f|^2,
\end{equation}
with the gravitino mass and the F-term energy of the visible fields
\begin{equation}\label{eq:m32Fvis}
m_{3/2}\equiv e^{k/2}|w|,\qquad |F_{\rm vis}|^2\equiv e^k\,k^{a\bar b}D_aw\,\overline{D_bw}.
\end{equation}
The negative term is the contribution of the compensator direction, and the two positive terms are those of the visible and goldstino directions. The goldstino auxiliary field has $|F^{\Xh}|=e^{k/2}|f|/h$ and carries the energy $U\equiv h|F^{\Xh}|^2=\mathcal{A}/h$. Wherever the gravitino mass dominates, $\mathcal{B}<0$, and we define a curvature scale $\LB$ by $\mathcal{B}=-3/\LB^2$.

\begin{figure}[t]
\centering
\includegraphics{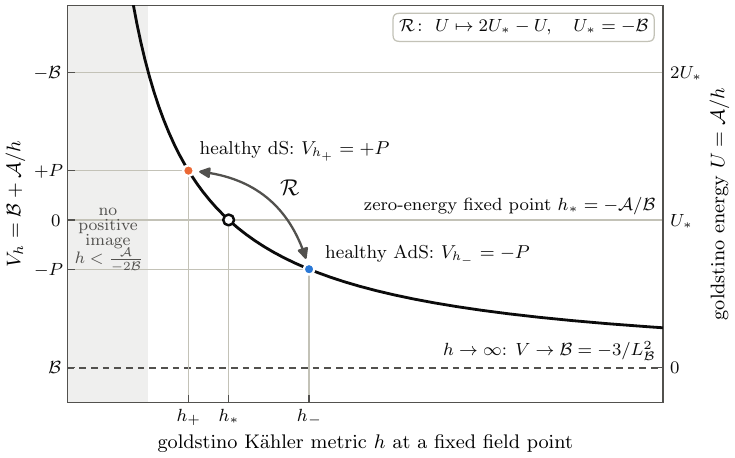}
\caption{The goldstino-metric reflection. The potential $V_h=\mathcal{B}+\mathcal{A}/h$ of Eq.~(\ref{eq:VAB}) is shown as a function of the goldstino K\"ahler metric $h$ at a fixed field point with $\mathcal{B}<0$, drawn for $\mathcal{B}=-3$, $\mathcal{A}=3$ and $P=1$ in units of $\LB$. The horizontal axis varies the goldstino metric $h$ at fixed scalar coordinates; it is not a scalar-field trajectory. The zero-energy fixed point $h_*=-\mathcal{A}/\mathcal{B}$ is a Minkowski vacuum only at points where the scalar stationarity conditions $\partial_aV=0$ also hold. The map $\RR$ exchanges the healthy dS member $h_+$, with $V=+P$, and the healthy AdS member $h_-=\RR(h_+)$, with $V=-P$, through $h_*$. Metrics with $h<\mathcal{A}/(-2\mathcal{B})$, for which $V_h>-\mathcal{B}$, have no positive image. The right axis shows the goldstino energy $U=\mathcal{A}/h=V_h-\mathcal{B}$, which vanishes as $h\to\infty$, where $V_h\to\mathcal{B}=-3/\LB^2$.}
\label{fig:reflection}
\end{figure}

The signs of section~\ref{sec:signs} act on the cone as a whole and can only relabel its directions. The weight $h$ instead changes the contribution of one positive direction, while the base $k$, the signs and the superpotential stay fixed. We work on an open region of field space on which $k_{a\bar b}$ is positive definite, $w$ and $f$ are holomorphic, $f$ has no zeros, and $h$ is smooth and positive, so that $\mathcal{A}>0$. By Eq.~(\ref{eq:VAB}) the weight controls the potential pointwise: any value $V>\mathcal{B}$ is obtained with $h=\mathcal{A}/(V-\mathcal{B})>0$. A metric $h'$ with the reversed potential must therefore satisfy
\begin{equation*}
V_{h'}=\mathcal{B}+\frac{\mathcal{A}}{h'}=-V_h\qquad\Longleftrightarrow\qquad\frac{\mathcal{A}}{h'}=-2\mathcal{B}-\frac{\mathcal{A}}{h}=\frac{-2h\mathcal{B}-\mathcal{A}}{h},
\end{equation*}
that is $U\mapsto-2\mathcal{B}-U$ for the goldstino energy. The image is positive and finite precisely when
\begin{equation*}
0<\frac{\mathcal{A}}{h}<-2\mathcal{B}\qquad\Longleftrightarrow\qquad\mathcal{B}<V_h<-\mathcal{B},
\end{equation*}
which requires $\mathcal{B}<0$ and defines the domain $|V_h|<-\mathcal{B}$, shown in figure~\ref{fig:reflection}. Solving for $h'$ gives the reflection
\begin{equation}\label{eq:R}
\boxed{\;\RR(h)=\frac{h\,\mathcal{A}}{-2h\,\mathcal{B}-\mathcal{A}},\qquad V_{\RR(h)}=-V_h,\qquad \RR\big(\RR(h)\big)=h,\;}
\end{equation}
where $\RR(\RR(h))=h$ follows from applying $U\mapsto-2\mathcal{B}-U$ twice; it maps positive metrics to positive metrics precisely on this domain. Equivalently, $\RR(h)=\mathcal{A}/(-V_h-\mathcal{B})$ is the metric that the reconstruction formula $h=\mathcal{A}/(V-\mathcal{B})$ of Refs.~\cite{Kallosh:2017wnt,McDonough:2016der} assigns to the reversed potential. Thus $\RR$ carries the reversal $V\mapsto-V$ of the potential over to the goldstino metric as a $\mathbb{Z}_2$ action, which relates two theories rather than reconstructing one; it acts at fixed $(k,w,f)$, and hence, as shown in section~\ref{sec:lagrangians}, at fixed visible kinetic metric and surviving unitary-gauge fermion operators. It acts on $h$ rather than on $f$, because $h$ is an arbitrary real function, whereas $|f|^2$ with holomorphic $f$ cannot reproduce a general real profile. Its fixed point $h_*\equiv-\mathcal{A}/\mathcal{B}$ is the metric at which the potential vanishes. A theory with $h=h_*$ everywhere has an identically vanishing potential and broken supersymmetry at every point, much as in no-scale supergravity \cite{Cremmer:1983bf}, and $\RR$ reflects the space of goldstino metrics through this flat theory: $h_*$ is its only fixed point, and it exchanges metrics with $V_h>0$ and $V_h<0$ point by point. Given a real function $P(z,\bar z)$ with $|P|<-\mathcal{B}$, which we call the target profile, the two members of a pair have the goldstino metrics
\begin{equation}\label{eq:hpm}
h_\pm=\frac{\mathcal{A}}{\pm P-\mathcal{B}}>0,\qquad V_{h_\pm}=\pm P,\qquad h_-=\RR(h_+).
\end{equation}

At fixed $(k,w,f)$ these are the only metrics with $V=\pm P$, and they are positive and finite precisely where $\mathcal{B}<-|P|$. Both members share $k$, $w$ and $f$, so all propagating scalars have the same positive kinetic metric $k_{a\bar b}$, and both goldstino metrics are positive. Where $P$ vanishes, the potentials and the values of the goldstino metrics coincide, $V_\pm=0$ and $h_+=h_-=h_*$, whereas the derivatives of $h_\pm$ differ by terms proportional to the derivatives of $P$, and the curvature of the potential is reversed: a Minkowski minimum of one member is a Minkowski maximum of the other. At either end of the domain, where $U\to0$ or $U\to-2\mathcal{B}$, one of the two members has a vanishing goldstino F-term and the constrained description degenerates, so the domain is open. Locally a pair exists for any profile. In a K\"ahler frame with $k=k_a=0$ at a point and constant $w$ with $3|w|^2>|P|$ there, one has $\mathcal{B}=-3|w|^2<-|P|$ at that point, and continuity gives a neighborhood with positive $h_\pm$.

\paragraph{Uplift interpretation.} In terms of the goldstino energy, $\RR$ is the reflection
\begin{equation}\label{eq:Ureflect}
U\ \mapsto\ 2U_*-U,\qquad U_*\equiv-\mathcal{B},
\end{equation}
about the value $U_*$ that cancels the background energy. With the uplift fraction $u\equiv U/U_*$ the potential is $V=-\mathcal{B}\,(u-1)$, the domain of $\RR$ is $0<u<2$, and $\RR$ acts as $u\mapsto2-u$. In string compactifications $U$ is the energy of an anti-D3-brane; we return to this interpretation in section~\ref{sec:strings}.

\paragraph{Covariance.} Under K\"ahler transformations combined with holomorphic rescalings of the goldstino superfield, $(k,w,\Xh,f)\to(k+\xi+\bar\xi,\,e^{-\xi}w,\,e^{\xi_X}\Xh,\,e^{-\xi-\xi_X}f)$, the quantities $\mathcal{B}$, $V_h$ and $\mathcal{A}/h$ are invariant, while $\mathcal{A}$, $h$ and $\RR(h)$ scale by the common factor $e^{-\xi_X-\bar\xi_X}$, so that critical points are mapped to critical points.

\subsection{The healthy dS and AdS Lagrangians}\label{sec:lagrangians}

The two members can now be written explicitly. On the healthy branch (\ref{eq:healthy}), and with $\tau=-\sigma_1$, the complete off-shell actions, including all fermions and auxiliary fields and before gauge fixing, are
\begin{empheq}[box=\fbox]{align}
\mathcal{L}_{\dS}&=\eL\Lsc\big[\mathcal{N}_{\sigma,h_+},\W\big]=\Lsc\Big[\Nstd\big[k+h_+\Xh\Xhb\big],\Big(\frac{X^0}{\sqrt3}\Big)^3\big(w+\Xh f\big)\Big],\label{eq:offshell_dS}\\[4pt]
\mathcal{L}_{\AdS}&=\eL\Lsc\big[\mathcal{N}_{\sigma,h_-},\W\big]=\Lsc\Big[\Nstd\big[k+h_-\Xh\Xhb\big],\Big(\frac{X^0}{\sqrt3}\Big)^3\big(w+\Xh f\big)\Big],\label{eq:offshell_AdS}
\end{empheq}
with $h_\pm$ given by Eq.~(\ref{eq:hpm}). They differ only in the weight of the goldstino direction of the cone.

\begin{table}[t]
\centering\small
\renewcommand{\arraystretch}{1.3}
\begin{tabular}{@{}lccll@{}}
\toprule
$(\eL,\sigma_1,\sigma_2)$ & $\eg$ & $K$ & kinetic terms & $V$\\
\midrule
\multicolumn{5}{@{}l}{\emph{$W=0$: goldstino direction inside the K\"ahler function, $h=\sigma_2$, and $w=0$}}\\
$(+,-,-)$, $(-,+,-)$ & $+1$ & $\K-\Xh\Xhb$ & ghosts: $\K_{a\bar b}<0$, $h<0$ & $-f^2$ (AdS)\\
$(-,-,-)$, $(+,+,-)$ & $-1$ & $\K-\Xh\Xhb$ & gravity-multiplet ghosts only & $-f^2$ (AdS)\\
$(+,+,+)$, $(-,-,+)$ & $-1$ & $-\K+\Xh\Xhb$ & all signs reversed & $+f^2$ (dS)\\
$(-,+,+)$, $(+,-,+)$ & $+1$ & $-\K+\Xh\Xhb$ & healthy & $+f^2$ (dS)\\
\midrule
\multicolumn{5}{@{}l}{\emph{$W\neq0$: healthy branch with an independent goldstino weight $h_\pm>0$}}\\
$(-,+,+)$, $(+,-,+)$ & $+1$ & $-\K+h_+\Xh\Xhb$ & healthy & $+P>0$ (dS)\\
$(-,+,+)$, $(+,-,+)$ & $+1$ & $-\K+h_-\Xh\Xhb$ & healthy, $h_-=\RR(h_+)$ & $-P<0$ (AdS)\\
\bottomrule
\end{tabular}
\caption{The two controls of the superconformal cone (\ref{eq:Nsigma}) for a K\"ahler function with negative-definite metric, $\K_{a\bar b}<0$. Upper block: the goldstino direction is part of the K\"ahler function, $\K\to\K-\Xh\Xhb$, which corresponds to $h=\sigma_2$, and $w=0$, so that $W=0$ on the bosonic locus; $V$ is the value of the potential at a point with $\K=0$, for real constant $f$, which is a vacuum where also $\partial_a\K=0$. Rows related by an overall sign have identical classical solutions, and only the healthy row is ghost-free; its potential is positive. Lower block: the healthy branch with an independent goldstino weight $h_\pm$ from Eq.~(\ref{eq:hpm}) and $w\neq0$; at a critical point of a target profile $P>0$ the values $V=\pm P$ are the cosmological constants of the two members. The two rows are the healthy Lagrangians (\ref{eq:L_dS}) and (\ref{eq:L_AdS}).}
\label{tab:signs}
\end{table}

For $F^{\Xh}\neq0$ the constraint $\Xh^2=0$ is solved by \cite{Rocek:1978nb,Komargodski:2009rz}
\begin{equation*}
\Xh=\frac{(\chi^{\Xh})^2}{2F^{\Xh}}+\sqrt2\,\theta\chi^{\Xh}+\theta^2F^{\Xh},
\end{equation*}
so that the scalar of $\Xh$ is a bilinear of the goldstino fermion $\chi^{\Xh}$. The goldstino auxiliary field is non-zero in both members, $F^{\Xh}\propto f\neq0$, so that the local supersymmetry transformation $\delta\chi^{\Xh}=\sqrt2\,F^{\Xh}\epsilon+\cdots$ can be used to set $\chi^{\Xh}=0$ in either member. In this unitary gauge the scalar $(\chi^{\Xh})^2/(2F^{\Xh})$ vanishes, while $F^{\Xh}$ survives \cite{Kallosh:2015tea,Hasegawa:2015bza,Schillo:2015ssx}. After superconformal gauge fixing and elimination of the visible auxiliary fields, the Einstein-frame Lagrangians are
\begin{equation}\label{eq:unitary}
e^{-1}\mathcal{L}_\pm=\frac12R-k_{a\bar b}\,\partial_\mu z^a\partial^\mu\bar z^{\bar b}-\mathcal{B}+h_\pm|F^{\Xh}|^2+\big(e^{k/2}fF^{\Xh}+\text{h.c.}\big)+e^{-1}\Lf[k,w].
\end{equation}
The goldstino auxiliary field enters only through the two terms shown (see the fourth fact below). At $\Xh=0$ one has $K_{\Xh\bar{\hat X}}=h_\pm$, $K_{a\bar{\hat X}}=0$ and $D_{\Xh}W=f$, so that $F^{\Xh}$ does not mix with the visible auxiliary fields $F^a$, whose elimination gives $-\mathcal{B}$. Completing the square,
\begin{equation*}
h_\pm|F^{\Xh}|^2+\big(e^{k/2}fF^{\Xh}+\text{h.c.}\big)=h_\pm\Big|F^{\Xh}+\frac{e^{k/2}\bar f}{h_\pm}\Big|^2-\frac{\mathcal{A}}{h_\pm},
\end{equation*}
the equation of motion $F^{\Xh}=-e^{k/2}\bar f/h_\pm$ leaves $-\mathcal{A}/h_\pm$, and the healthy dS and AdS Lagrangians are
\begin{empheq}[box=\fbox]{align}
e^{-1}\mathcal{L}_{\dS}&=\frac12R-k_{a\bar b}\,\partial_\mu z^a\partial^\mu\bar z^{\bar b}-\Big(\mathcal{B}+\frac{\mathcal{A}}{h_+}\Big)+e^{-1}\Lf[k,w],\qquad V_{\dS}=+P,\label{eq:L_dS}\\[4pt]
e^{-1}\mathcal{L}_{\AdS}&=\frac12R-k_{a\bar b}\,\partial_\mu z^a\partial^\mu\bar z^{\bar b}-\Big(\mathcal{B}+\frac{\mathcal{A}}{h_-}\Big)+e^{-1}\Lf[k,w],\qquad V_{\AdS}=-P.\label{eq:L_AdS}
\end{empheq}
The fermionic Lagrangian is that of standard supergravity with K\"ahler potential $k$ and superpotential $w$. In the conventions of Ref.~\cite{Freedman:2012zz}, eqs.~(18.6)--(18.19), it reads
\begin{align}
e^{-1}\Lf[k,w]=&-\frac12\bar\psi_\mu\gamma^{\mu\nu\rho}D_\nu\psi_\rho-\frac12k_{a\bar b}\big(\bar\chi^aD\!\!\!/\,\chi^{\bar b}+\bar\chi^{\bar b}D\!\!\!/\,\chi^a\big)\nonumber\\
&+\Big(\frac{1}{\sqrt2}k_{a\bar b}\,\bar\psi_\mu\,\partial\!\!\!/\bar z^{\bar b}\gamma^\mu\chi^a+\frac12e^{k/2}w\,\bar\psi_\mu P_R\gamma^{\mu\nu}\psi_\nu\nonumber\\
&\qquad+\frac{1}{\sqrt2}e^{k/2}D_aw\,\bar\psi_\mu\gamma^\mu\chi^a-\frac12m_{ab}\,\bar\chi^a\chi^b+\text{h.c.}\Big)+e^{-1}\mathcal{L}_{4f}.\label{eq:Lferm}
\end{align}
Here $\psi_\mu$ is the Majorana gravitino, $\chi^a=P_L\chi^a$ are the visible chiral fermions and $\chi^{\bar b}=P_R\chi^{\bar b}$ their charge conjugates, with $P_{L,R}=\frac12(1\pm\gamma_*)$ and $\gamma_*=i\gamma_0\gamma_1\gamma_2\gamma_3$; a bar on a spinor denotes its Majorana conjugate, $\bar\lambda=\lambda^TC$ \cite{Freedman:2012zz}. The matrices $\gamma^{\mu\nu}$ and $\gamma^{\mu\nu\rho}$ are antisymmetrized products of gamma matrices with unit weight, $D\!\!\!/=\gamma^\mu D_\mu$ and $\partial\!\!\!/=\gamma^\mu\partial_\mu$. The covariant derivative $D_\mu$ contains the spin connection, the composite $U(1)$ connection built from $k_a\partial_\mu z^a-k_{\bar a}\partial_\mu\bar z^{\bar a}$, with $k_{\bar a}\equiv\partial_{\bar a}k$, and, acting on $\chi^a$, the Christoffel symbols $\Gamma^c_{ab}\equiv k^{c\bar d}\partial_ak_{b\bar d}$ of the visible metric. The fermion mass matrix is
\begin{equation}\label{eq:mab}
m_{ab}=e^{k/2}\nabla_aD_bw,\qquad \nabla_aD_bw\equiv\partial_aD_bw+k_aD_bw-\Gamma^c_{ab}D_cw,
\end{equation}
and $\mathcal{L}_{4f}$ denotes the four-fermion terms, which depend on $k$ only through $k_{a\bar b}$ and its curvature \cite{Freedman:2012zz}. Four facts make Eq.~(\ref{eq:Lferm}) independent of $h$. First, in the block-diagonal class $K_{\Xh}$, $K_{a\bar{\hat X}}$ and $\Gamma^{\Xh}_{ab}$ vanish at $\Xh=0$, so that the Christoffel symbols $\Gamma^c_{ab}$, the composite $U(1)$ connection and the mass matrix (\ref{eq:mab}) are those of $k$. Second, the curvature in $\mathcal{L}_{4f}$ is that of $k$: since $K_{a\bar{\hat X}}=\Xh\,\partial_ah$ vanishes at $\Xh=0$,
\begin{equation*}
R_{a\bar bc\bar d}[K]=\partial_c\partial_{\bar d}K_{a\bar b}-K^{i\bar j}\,\partial_cK_{a\bar j}\,\partial_{\bar d}K_{i\bar b}\qquad\Longrightarrow\qquad R_{a\bar bc\bar d}[K]\big|_{\Xh=0}=R_{a\bar bc\bar d}[k].
\end{equation*}
Third, the gravitino mass $m_{3/2}=e^{K/2}|W|=e^{k/2}|w|$ is common to both members. Fourth, every fermionic term that contains $F^{\Xh}$ or the connection component $\Gamma^{\Xh}_{a\Xh}=\partial_a\ln h$, which need not vanish, also contains $\chi^{\Xh}$ and vanishes in unitary gauge \cite{Kallosh:2015tea,Hasegawa:2015bza,Schillo:2015ssx}.

Equations (\ref{eq:offshell_dS})--(\ref{eq:Lferm}) are the central result of this paper. The two Lagrangians have the same field content, the same positive visible kinetic metric $k_{a\bar b}>0$, positive goldstino metrics $h_\pm>0$, identical surviving unitary-gauge ($\chi^{\Xh}=0$) fermion operators $\Lf[k,w]$ and opposite potentials. The fermionic Lagrangians coincide as functionals of the fields; on the different backgrounds of the two members they give different spectra. With visible supersymmetry breaking, $D_aw\neq0$, the unitary gauge removes only the fermion of the nilpotent multiplet, and the gravitino--visible fermion mixing is retained in both members; with vector multiplets, a $\Xh$-dependent gauge kinetic function would require a separate check. Table~\ref{tab:signs} summarizes the two layers of the construction.

\section{The dS/AdS Dictionary}\label{sec:dictionary}

\subsection{Vacua and spectra}\label{sec:vacua}

Let $z_0$ be a critical point of the target profile $P$ with $P(z_0)>0$. By Eqs.~(\ref{eq:L_dS})--(\ref{eq:L_AdS}) it is a critical point of both members: a dS vacuum of the dS member, with $V=3H^2=P(z_0)$, and an AdS vacuum of the AdS member, with $V=-3/L^2=-P(z_0)$. Hence
\begin{equation}\label{eq:HL}
H=\frac1L.
\end{equation}
Since the kinetic metric is shared and the potentials are opposite, the mass matrices of all scalar fluctuations are opposite,
\begin{equation}\label{eq:masses}
m^2_{\dS}=-m^2_{\AdS},
\end{equation}
where $m^2$ denotes the eigenvalues of the Hessian of the potential at $z_0$ in a basis of canonically normalized real fields.

Every critical point of one member within the domain of $\RR$ is therefore a critical point of the other, with the opposite Hessian, and the character of the critical point is reversed:
\begin{itemize}
\item A dS minimum, with $V_+>0$ and $m^2_{\dS}>0$ in all directions, is paired with an AdS maximum, with $V_-<0$ and $m^2_{\AdS}<0$. In AdS$_4$ a local maximum of the potential is a linearly stable vacuum as long as its curvature obeys the BF bound $m^2_{\AdS}\ge-9/(4L^2)$ \cite{Breitenlohner:1982jf}; this is the window $0<m^2_{\dS}\le\frac94H^2$ of the complementary series discussed in section~\ref{sec:window}.
\item A dS maximum, or hilltop, with $m^2_{\dS}<0$, is paired with a genuine AdS minimum, with $m^2_{\AdS}>0$, which is always stable. A hilltop with $|m^2_{\dS}|\ll H^2$ is the starting region of hilltop slow-roll inflation, so that inflationary hilltops of the dS member are paired with AdS minima of its partner.
\item At a zero of $P$, a Minkowski minimum of one member is paired with a Minkowski maximum of the other. In Minkowski space there is no BF protection, so the reflection of an isolated stable Minkowski minimum is unstable, and the two members cannot both have an isolated stable Minkowski minimum at the same point.
\end{itemize}
Saddle points are paired with saddle points, with every Hessian eigenvalue reversed. Section~\ref{sec:multivacuum} constructs a closed-form inflationary pair that realizes all three pairings.

A scalar of mass $m$ has the conformal weights \cite{Strominger:2001pn}
\begin{equation}\label{eq:weights}
\Delta^{\dS}_\pm=\frac32\pm\sqrt{\frac94-\frac{m^2_{\dS}}{H^2}},\qquad \Delta^{\AdS}_\pm=\frac32\pm\sqrt{\frac94+m^2_{\AdS}L^2},
\end{equation}
which govern its late-time falloff in dS$_4$ and its falloff near the boundary of AdS$_4$. With Eqs.~(\ref{eq:HL})--(\ref{eq:masses}) the two coincide,
\begin{equation}\label{eq:Delta}
\Delta_{\dS}=\Delta_{\AdS}.
\end{equation}
This is an identity between the indicial roots of the linearized field equations; the admissible AdS boundary conditions and the dS state are additional data.

\subsection{Stability window}\label{sec:window}

The stability conditions of the two members are two sides of the same bound. With standard reflecting boundary conditions, the AdS member is linearly stable if every scalar satisfies the BF bound $m^2_{\AdS}\ge-9/(4L^2)$ \cite{Breitenlohner:1982jf}. By Eq.~(\ref{eq:masses}) this is the condition $m^2_{\dS}\le\frac94H^2$, which states that the corresponding dS fields lie in the complementary series of the de Sitter group \cite{Sun:2021thf}, where the weights (\ref{eq:Delta}) are real. A tachyon-free dS vacuum therefore has a stable AdS partner if and only if every scalar mode lies in the window
\begin{equation}\label{eq:window}
0\le\eta\le\frac34,\qquad \eta\equiv\frac{m^2_{\dS}}{3H^2}=-\frac{m^2_{\AdS}L^2}{3},
\end{equation}
which is shown in figure~\ref{fig:window}. Equivalently, both members are linearly stable precisely when all weights are real and $\Delta_-\ge0$; a dS tachyon has real weights with $\Delta_-<0$. The endpoints are limiting cases: a massless minimally coupled scalar in dS has infrared subtleties, and at BF saturation the AdS falloffs require a choice of boundary condition. The reflected inequality $m^2\le\frac94H^2$ was noted as a cosmological analogue of the BF bound in Ref.\cite{Skenderis:2006rr,Skenderis:2006fb}; here it becomes a two-sided statement about two healthy theories. The argument uses only the shared kinetic metric and the opposite potentials, and applies to any pair of theories with these two properties. A dS vacuum with a scalar heavier than $\frac32H$ has a BF-unstable partner.

The weights also govern the asymptotic linear relaxation of the dS member in the quadratic basin of the vacuum, where the homogeneous mode of a scalar obeys $\delta\ddot\phi+3H\delta\dot\phi+m^2\delta\phi=0$ and is a combination of $e^{-\Delta_+Ht}$ and $e^{-\Delta_-Ht}$. In the window this relaxation is overdamped, while a heavier scalar is underdamped and oscillates with the envelope $e^{-3Ht/2}$, as noted for this cosmological bound in Refs.~\cite{Skenderis:2006rr,Skenderis:2006fb}; at the critical value $m^2=\frac94H^2$ the solution is $(c_1+c_2t)\,e^{-3Ht/2}$, and at $m^2=0$ a constant marginal mode remains. These statements concern the linearized dynamics: outside the quadratic basin, anharmonic terms can produce transient oscillations before the linear regime is reached. In section~\ref{sec:holography} this threshold is identified with the onset of oscillatory squeezed-limit signals, in section~\ref{sec:strings} it separates two classes of string-derived dS vacua, and in section~\ref{sec:inflation} it decides whether the final linear relaxation of an inflaton is oscillatory.

\subsection{Gravitino}

Both members have the same gravitino mass $m_{3/2}=e^{k/2}|w|$. For a massive spin-3/2 field in (A)dS$_4$ with cosmological constant $\Lambda$ the unitary region is $m_{3/2}^2\ge-\Lambda/3$, the boundary being the gauge-invariant line of the massless gravitino \cite{Deser:2001pe}. In dS$_4$ this allows every real mass, unlike the Higuchi bound $m^2\ge2H^2$ for massive spin two \cite{Higuchi:1986py}; in AdS$_4$ it requires $m_{3/2}\ge1/L$. Any configuration of a healthy supergravity obeys $V=|F|^2-3m_{3/2}^2$, where $|F|^2\ge0$ is the total F-term energy. For the AdS member, with $V=-3/L^2$ and $|F_{\AdS}|^2=|F_{\rm vis}|^2+\mathcal{A}/h_-$, this gives
\begin{equation}\label{eq:m32AdS}
m_{3/2}^2=\frac{1}{L^2}+\frac13|F_{\AdS}|^2\ \ge\ \frac1{L^2},
\end{equation}
and the AdS member satisfies the bound automatically and saturates it only where it is supersymmetric. Because the gravitino mass is shared and $H=1/L$, the dS member inherits the reflected bound
\begin{equation}\label{eq:m32dS}
m_{3/2}^2\ \ge\ \frac{\Lambda_{\dS}}{3}=H^2,\qquad\text{and}\qquad m_{3/2}^2>H^2+\frac13|F_{\rm vis}|^2\quad\text{within the domain of }\RR.
\end{equation}
For the dS member this is not a unitarity requirement. It has the form of the bound $H\lesssim m_{3/2}$ found for KKLT in Ref.~\cite{Kallosh:2004yh}, but here it is enforced by the existence of a ghost-free AdS partner. Within the goldstino mechanism pairs exist only where $3m_{3/2}^2>|P|+|F_{\rm vis}|^2$, and the AdS member is never supersymmetric at finite $h_-$. Without supersymmetry breaking in the goldstino direction, $f=0$, the potential does not depend on $h$ and no pair exists. The value $m_{3/2}=H$ is approached only when the AdS member becomes supersymmetric, which requires $F_{\rm vis}=0$ and the decoupling limit $h_-\to\infty$, a boundary of the regular nilpotent description rather than a member of a pair.

The reflected bound also shows how the correspondence is consistent with the absence of unitary de Sitter superalgebras \cite{Pilch:1984aw,Lukierski:1984it}. The supersymmetric line $m_{3/2}=1/L$ of the AdS member is mapped onto $m_{3/2}=H$, which in dS$_4$ is an ordinary massive gravitino, whereas the gauge-invariant line of dS$_4$, where an unbroken de Sitter supersymmetry would have to reside, lies at $m_{3/2}^2=-H^2$. A supersymmetric AdS vacuum is therefore never mapped onto a supersymmetric dS vacuum, and, as in pure de Sitter supergravity \cite{Bergshoeff:2015tra}, the dS members break supersymmetry spontaneously.

\section{Holographic Cosmology, dS/CFT, and Cosmological Correlators}\label{sec:holography}

The reversal of the potential acts on solutions as well as on vacua. Consider the bosonic Einstein-frame action with kinetic metric $k_{a\bar b}$ and potential $V$, and the two ans\"atze
\begin{equation}\label{eq:ansatz}
ds^2_{\text{cos}}=-N(t)^2dt^2+e^{2\rho(t)}d\vec x^{\,2},\qquad ds^2_{\text{wall}}=N(r)^2dr^2+e^{2\rho(r)}\big(-dx_0^2+dx_1^2+dx_2^2\big),
\end{equation}
with scalar profiles $z^a(t)$ or $z^a(r)$, a lapse function $N$ and a scale factor $e^{\rho}$. After removing total derivatives, which are cancelled by the Gibbons--Hawking--York term, the reduced Lagrangians per unit volume of the slices are
\begin{align}
\mathcal{L}_{\text{cos}}[V]&=e^{3\rho}\Big(-\frac{3\dot\rho^2}{N}+\frac{k_{a\bar b}\dot z^a\dot{\bar z}^{\bar b}}{N}-NV\Big),\label{eq:Lcos}\\
\mathcal{L}_{\text{wall}}[V]&=e^{3\rho}\Big(+\frac{3\rho'^2}{N}-\frac{k_{a\bar b}z'^a\bar z'^{\bar b}}{N}-NV\Big),\label{eq:Lwall}
\end{align}
where a dot and a prime denote derivatives with respect to $t$ and $r$. For the two members of a pair $V_{h_+}=-V_{h_-}$, and therefore
\begin{equation}\label{eq:LcosLwall}
\mathcal{L}_{\text{cos}}[V_{h_+}](t)=-\mathcal{L}_{\text{wall}}[V_{h_-}](r\to t).
\end{equation}
The identity holds off shell in the reduced fields, including the lapse. Every flat domain wall of the AdS member that remains in the common domain of $\RR$ is therefore a flat FLRW cosmology of the dS member, with the same scalar and scale-factor profiles, and vice versa; variation with respect to the lapse maps the two Hamiltonian constraints onto each other. In particular, an inflationary trajectory of the dS member is a domain wall of the AdS member, which in a holographic setting describes a renormalization-group flow \cite{McFadden:2009fg}.

This is the domain-wall/cosmology correspondence of Refs.~\cite{Skenderis:2006jq,Skenderis:2007sm}. In its supersymmetric form \cite{Skenderis:2007sm} the cosmological partner of a domain wall of $\mathcal{N}=2$ gauged supergravity is a pseudo-supergravity, with twisted reality conditions on the spinors and an imaginary gauge coupling; such theories contain vector ghosts and are not unitary \cite{Skenderis:2007sm,Bergshoeff:2007cg}, and related constructions use variant supergravities with supersymmetric dS vacua \cite{Vaula:2007jk}. In the pairs constructed here both members are ordinary $\mathcal{N}=1$ supergravities with the same field content, visible kinetic terms and surviving fermion operators (section~\ref{sec:lagrangians}). What makes this possible is that the reversal is carried by a coupling of the theory, the goldstino metric, rather than by a continuation of the fields or of the gauge coupling. The price is that supersymmetry is broken in both members, so that the walls and cosmologies preserve none of the supersymmetries of the two theories. On monotonic single-scalar segments the wall and the cosmology are still governed by a common local fake superpotential \cite{Skenderis:2006jq,Skenderis:2006fb}, a structure of the bosonic equations that does not rely on the supersymmetries of the theory.

In holographic cosmology \cite{McFadden:2009fg} the correspondence is implemented by continuing the gravitational coupling $\kappa^2=M_{\rm Pl}^{-2}$ to $-\kappa^2$ and the momenta $p$ to $-ip$, equivalently by a continuation of the rank of the gauge group of the dual field theory. In Eq.~(\ref{eq:LcosLwall}) the overall sign is precisely the continuation of $\kappa^2$, while the potentials of the two members are related by $V\to-V$. In the continuation that relates the wavefunction of the universe to the AdS partition function \cite{Maldacena:2002vr}, the potential is reversed by continuing the curvature radius. Here it is reversed within real field space by the goldstino-metric reflection of section~\ref{sec:reflection}, and only the coordinates and the overall sign of the reduced action are continued. The two members thus provide two healthy bulk actions on which such a holographic relation can be studied; establishing it requires, in addition, a contour and state prescription, the fluctuations and the boundary counterterms.

The same pairing organizes the conformal weights. In dS/CFT a bulk scalar of mass $m$ is associated with the weights $\Delta_\pm$ of section~\ref{sec:dictionary}, which become complex for $m>\frac32H$ \cite{Strominger:2001pn}. Equation~(\ref{eq:Delta}) identifies them with the scaling dimensions of the AdS member at $L=1/H$, and the window (\ref{eq:window}) states that a tachyon-free pair is stable exactly when all weights are real. During inflation the same weights control the squeezed limit of the bispectrum of the comoving curvature perturbation $\zeta$ \cite{Chen:2009we,Chen:2009zp,Arkani-Hamed:2015bza}. A weakly coupled spectator field of approximately constant mass $m$ contributes, for suitable couplings and the usual initial state, a non-analytic term
\begin{equation}\label{eq:squeezed}
\frac{\langle\zeta_{\mathbf p_L}\zeta_{\mathbf p_S}\zeta_{-\mathbf p_L-\mathbf p_S}\rangle'}{P_\zeta(p_L)P_\zeta(p_S)}\ \supset\
\begin{cases}
C\,(p_L/p_S)^{\Delta_-}, & 0<m^2<\frac94H^2,\\[4pt]
C'\,(p_L/p_S)^{3/2}\cos\!\big(\nu\ln(p_L/p_S)+\delta\big), & m^2=\big(\frac94+\nu^2\big)H^2,
\end{cases}
\end{equation}
with $\nu>0$, where $p_L\ll p_S$ are the long and short momenta, $P_\zeta$ is the power spectrum, the prime denotes omission of the momentum-conserving delta function, and $C$, $C'$ and $\delta$ depend on the interactions. Fields in the complementary series can give power laws, and fields in the principal series, whose weights $\frac32\pm i\nu$ are complex, can give the oscillations by which inflation acts as a cosmological collider \cite{Arkani-Hamed:2015bza}. Through Eq.~(\ref{eq:Delta}) the squeezed-limit exponents of the dS member are the scaling dimensions of its AdS partner, and an oscillatory signal is the dS image of a complex scaling dimension, the hallmark of a BF violation in AdS. At paired vacua a stable AdS partner therefore exists only for spectra without principal-series fields. The amplitudes $C$ and $C'$ can vanish, and oscillations can also have other origins; the weight identity is exact at paired vacua, while its use for observables of a rolling background requires the full fluctuation calculation, in which mixing and field-space curvature also enter.

The relation between the two members also has a covariant counterpart. In four dimensions the Ricci scalar of the metric $-g_{\mu\nu}$ is $-R$, while $\sqrt{|\det g_{\mu\nu}|}$ is invariant, so that the action $S[g;k,V]$ of the bosonic theory obeys
\begin{equation}\label{eq:sigrev}
S[g;k,V]=-S[-g;k,-V].
\end{equation}
For homogeneous configurations the signature of the flat slices does not enter, and Eq.~(\ref{eq:LcosLwall}) is the reduced form of this identity. Neither identity defines a real isometry between AdS and dS; the wall/cosmology map also exchanges the causal role of the slices. The same symmetry underlies the accelerating histories found in quantum cosmology with a negative cosmological constant \cite{Hartle:2012qb,Mithani:2013ed}. There both continuations live in a single theory, and the inflating solutions generally develop tachyonic and ghost instabilities \cite{Mithani:2013ed}. Here the accelerating and the AdS histories belong to two different healthy theories, related by the goldstino-metric reflection.

\begin{figure}[t]
\centering
\includegraphics{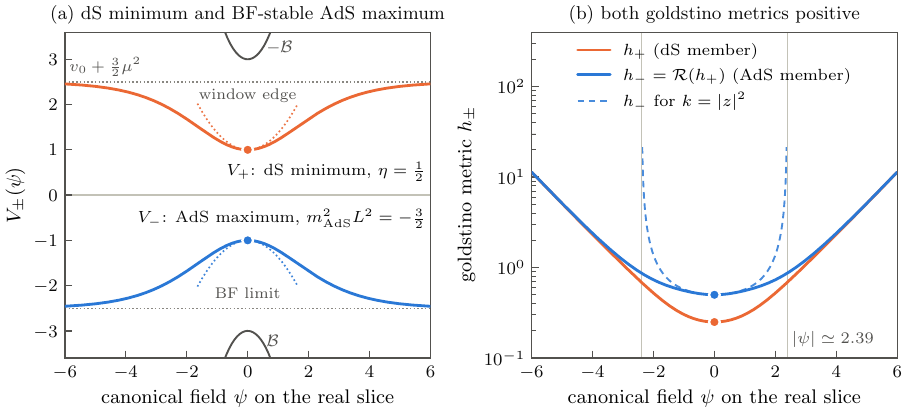}
\caption{The explicit pair (\ref{eq:model}) at the benchmark $w_0=f_0=v_0=\mu^2=1$ along the real slice, with the canonical field $\psi$ of Eq.~(\ref{eq:psi}). (a) The potentials $V_\pm=\pm[1+\frac32\tanh^2(\psi/\sqrt6)]$. The dS member has a minimum at $\psi=0$ with $\eta=\frac12$, inside the window, and the AdS member a local maximum with $m^2_{\AdS}L^2=-\frac32>-\frac94$, a linearly BF-stable AdS vacuum; the dotted parabolas show the curvatures $m^2=\pm\frac94H^2$ at the edge of the window and at the BF bound. Both potentials approach $\pm(v_0+\frac32\mu^2)$ as $\psi\to\pm\infty$, and the gray curves $\pm\mathcal{B}=\mp3\cosh^4(\psi/\sqrt6)$ bound the domain $|V|<-\mathcal{B}$ of $\RR$. (b) The goldstino metrics $h_\pm$ of Eq.~(\ref{eq:hpm_model}), positive on the whole slice, with $h_+(0)=\frac14$, $h_-(0)=\frac12$ and $h_\pm\simeq\frac13\cosh^2(\psi/\sqrt6)$ at large $|\psi|$. The dashed curve is $h_-$ for the flat K\"ahler potential $k=|z|^2$ with the same $w$, $f$ and $P$, plotted against its canonical field $\psi=\sqrt2\,\text{Re}\,z$; it diverges at $|\psi|\simeq2.39$ and is negative beyond.}
\label{fig:disk}
\end{figure}

\section{An Explicit dS/AdS Pair}\label{sec:example}

Take one visible chiral multiplet with
\begin{equation}\label{eq:model}
k=-3\ln\Big(1-\frac{|z|^2}{3}\Big),\qquad w=w_0,\qquad f=f_0,\qquad P=v_0+\frac{\mu^2}{2}|z|^2,
\end{equation}
with positive constants $w_0$, $f_0$, $v_0$ and $\mu^2$. In superconformal language this is the canonical cone. With $\sigma_1=\sigma_2=+1$ and $\K=-k$, Eq.~(\ref{eq:Nsigma}) gives
\begin{equation}\label{eq:cone_model}
\mathcal{N}_{\sigma,h}=\big(|X^0|^2-|X^1|^2\big)\Big(1-\frac h3\,\Xh\Xhb\Big),
\end{equation}
and $\eL=-1$ turns $|X^0|^2-|X^1|^2$ into the canonical cone potential $-|X^0|^2+|X^1|^2$ \cite{Ferrara:2010in}. The field space is the hyperbolic disk $|z|^2<3$, on which
\begin{equation}\label{eq:AB_model}
\mathcal{A}=\frac{f_0^2}{(1-|z|^2/3)^3},\qquad \mathcal{B}=-\frac{3w_0^2}{(1-|z|^2/3)^2}\le-3w_0^2 .
\end{equation}
The background energy is negative everywhere and diverges at the boundary. A simple sufficient condition for $|P|<-\mathcal{B}$ on the whole field space is
\begin{equation}\label{eq:global}
v_0+\frac32\mu^2<3w_0^2,
\end{equation}
which is sharpened below. Under it the two goldstino metrics
\begin{equation}\label{eq:hpm_model}
h_\pm=\frac{f_0^2\,(1-|z|^2/3)^{-3}}{\pm\big(v_0+\frac12\mu^2|z|^2\big)+3w_0^2(1-|z|^2/3)^{-2}}
\end{equation}
are positive everywhere, and the two members have the potentials $V_\pm=\pm P$ exactly. The model is invariant under $z\to\bar z$, so that the real slice $\text{Im}\,z=0$ is a consistent truncation. Along it the canonically normalized field $\psi$, with kinetic term $-\frac12(\partial\psi)^2$, is defined by
\begin{equation}\label{eq:psi}
z=\sqrt3\tanh\frac{\psi}{\sqrt6},\qquad V_\pm(\psi)=\pm\Big[v_0+\frac32\mu^2\tanh^2\frac{\psi}{\sqrt6}\Big],
\end{equation}
shown in figure~\ref{fig:disk}(a). The dS member has a minimum at $\psi=0$, the AdS member is its mirror image with a maximum there, and both approach $\pm(v_0+\frac32\mu^2)$ at the boundary of the disk, $\psi\to\pm\infty$. With the flat K\"ahler potential $k=|z|^2$ instead, the background energy turns positive at $|z|^2=3$, and at the benchmark below the pair exists only on a disk around the origin, beyond which $h_-$ diverges and turns negative (figure~\ref{fig:disk}(b)).

At $z=0$ one has $k_{z\bar z}=1$, $D_zw=0$ and $\nabla_zD_zw=0$. In terms of the real fields $\phi_1,\phi_2$, defined by $z=(\phi_1+i\phi_2)/\sqrt2$ and canonically normalized at $z=0$, the two vacua are
\begin{equation}\label{eq:vacua_model}
\begin{aligned}
&\text{dS member:}&\qquad V&=+v_0=3H^2,&\qquad m^2_{\phi_{1,2}}&=+\frac{\mu^2}{2},\\
&\text{AdS member:}&\qquad V&=-v_0=-\frac{3}{L^2},&\qquad m^2_{\phi_{1,2}}&=-\frac{\mu^2}{2},
\end{aligned}
\end{equation}
with the common gravitino mass $m_{3/2}=w_0$ and a massless visible fermion. The AdS scalars obey the BF bound when $\mu^2\le\frac32v_0$, which is the window (\ref{eq:window}), and the gravitino satisfies $m_{3/2}^2L^2=m_{3/2}^2/H^2=3w_0^2/v_0>1$, as required by Eqs.~(\ref{eq:m32AdS})--(\ref{eq:m32dS}). For the benchmark $w_0=f_0=v_0=\mu^2=1$ one finds
\begin{equation}\label{eq:benchmark}
h_+(0)=\frac14,\qquad h_-(0)=\frac12,\qquad H^2=L^{-2}=\frac13,\qquad \eta=\frac12,\qquad m_{3/2}^2=3H^2,
\end{equation}
and the global condition (\ref{eq:global}) reads $\frac52<3$. Both members are ghost-free on the entire field space. The dS vacuum is a minimum, and the AdS vacuum is a local maximum of $V_-$ that obeys the BF bound, $m^2_{\AdS}L^2=-\frac32>-\frac94$, and is therefore linearly stable.

The condition (\ref{eq:global}) is sufficient but not necessary. With $y\equiv|z|^2$, positivity of both metrics on the whole disk is equivalent to
\begin{equation}\label{eq:sharpdisk}
3w_0^2>\max_{0\le y<3}\Big(v_0+\frac{\mu^2}{2}\,y\Big)\Big(1-\frac y3\Big)^2=\begin{cases}
v_0, & \mu^2\le\frac43v_0,\\[3pt]
\dfrac{2(2v_0+3\mu^2)^3}{243\,\mu^4}, & \mu^2>\frac43v_0,
\end{cases}
\end{equation}
the interior maximum lying at $y=1-4v_0/(3\mu^2)$; the benchmark satisfies it with a wide margin, $3>1$. Rescaling $(w_0,f_0)\to\lambda(w_0,f_0)$ and $(v_0,\mu^2)\to\lambda^2(v_0,\mu^2)$ with a positive constant $\lambda$ leaves $h_\pm$ and all dimensionless vacuum data unchanged. The supersymmetry-breaking scale $(\mathcal{A}/h)^{1/4}\propto\lambda^{1/2}$ then exceeds the curvature and mass scales, which are proportional to $\lambda$, for $\lambda\ll1$, as the nilpotent description requires. Positivity on the open disk is not the same as uniform effective-field-theory control near its boundary, which lies at infinite distance.

\section{String Compactifications, Anti-D3 Uplifting, and the Moduli Mass Dichotomy}\label{sec:strings}

\subsection{Flux backgrounds and the anti-D3-brane}

In type IIB flux compactifications the three-form fluxes stabilize the complex-structure moduli and the dilaton and leave, at tree level, a no-scale Minkowski vacuum for the K\"ahler moduli \cite{Giddings:2001yu}. Non-perturbative effects stabilize the K\"ahler moduli in supersymmetric AdS vacua \cite{Kachru:2003aw}, and together with $\alpha'$ corrections they can lead to non-supersymmetric AdS vacua at exponentially large volume \cite{Balasubramanian:2005zx,Conlon:2005ki}. AdS and Minkowski vacua are thus abundant and comparatively well controlled. De Sitter vacua, by contrast, are excluded in classical two-derivative compactifications that satisfy the assumptions of the Maldacena--Nu\~nez theorem \cite{Maldacena:2000mw}. The best-studied proposals add an anti-D3-brane at the tip of a warped throat \cite{Kachru:2003aw}, whose metastability against brane/flux annihilation was analyzed in Ref.~\cite{Kachru:2002gs}, and their consistency remains under discussion \cite{Danielsson:2018ztv,Obied:2018sgi,Ooguri:2018wrx}.

In the four-dimensional effective theory the anti-D3-brane is described by a nilpotent goldstino multiplet \cite{Kallosh:2014wsa,Bergshoeff:2015jxa,Ferrara:2014kva,GarciadelMoral:2017vnz}, so that the potential has precisely the form (\ref{eq:VAB}),
\begin{equation}\label{eq:Vstring}
V=\mathcal{B}+U,\qquad \mathcal{B}=|F_{\rm vis}|^2-3m_{3/2}^2\equiv-\frac{3}{\LB^2},\qquad U=\frac{\mathcal{A}}{h}=h|F^{\Xh}|^2 .
\end{equation}
The background energy $\mathcal{B}$ is determined by the K\"ahler potential $k$ of the bulk moduli and by the flux and non-perturbative superpotential $w$. At a supersymmetric AdS vacuum it equals the vacuum energy $-3m_{3/2}^2$, and wherever it is negative it defines a curvature scale $\LB^{-1}\le m_{3/2}$, which is an actual AdS radius only at a constant-scalar solution of the background theory. The uplift $U$ is the energy of the anti-brane, and the goldstino metric encodes its dependence on the bulk moduli \cite{GarciadelMoral:2017vnz,McDonough:2016der,Kallosh:2017wnt}; an example is given in section~\ref{sec:kklt}.

\subsection{Over- and under-uplifting}

In this language $\RR$ is the reflection (\ref{eq:Ureflect}) of the uplift energy about its Minkowski value $U_*=-\mathcal{B}=3/\LB^2$. The self-dual theory $U=U_*$ is a four-dimensional analogue of the no-scale Minkowski vacuum, in which the anti-brane energy cancels the background energy at every point of moduli space. For $P>0$ the dS member is over-uplifted, $U_+=U_*+P$ with $h_+<h_*$, and the AdS member is under-uplifted by the same amount, $U_-=U_*-P$ with $h_->h_*$.

The AdS member is not the original model with a weaker anti-brane. Since $U_*$ depends on the moduli, its goldstino metric $h_-=\RR(h_+)$ is a different function of the moduli, which reverses the whole potential, including its curvature. As a four-dimensional effective theory it has the same functions $k$, $w$ and $f$ as the dS member and a non-supersymmetric AdS critical point, while its anti-brane energy changes from $U_+$ to $U_-$. Whether a goldstino metric of this form arises from a microscopic configuration, with compatible flux quantization, tadpole cancellation and controlled cutoffs for both metrics, is a question we leave open. The partner exists precisely when the dS vacuum lies in the domain of $\RR$, $H<1/\LB$, which by Eq.~(\ref{eq:m32dS}) implies $H<m_{3/2}$, the form of the bound found for KKLT in Ref.~\cite{Kallosh:2004yh}.

\begin{figure}[t]
\centering
\includegraphics{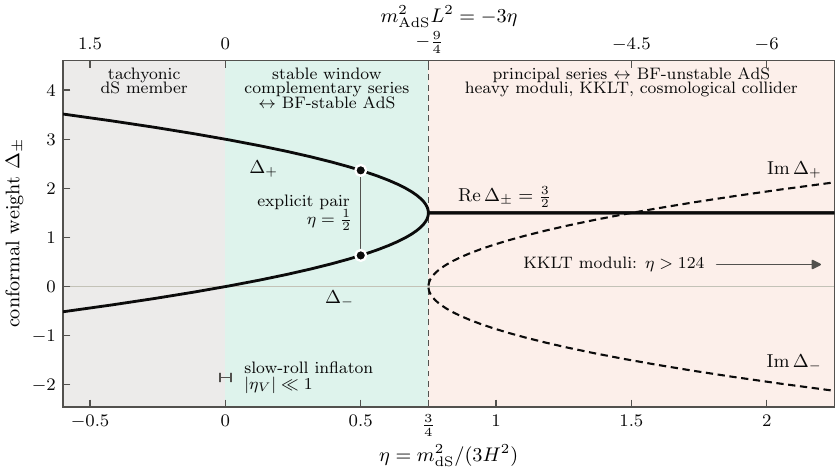}
\caption{Stability window and moduli mass dichotomy. The conformal weights $\Delta_\pm=\frac32\pm\sqrt{\frac94-3\eta}$ of a scalar mode are shown as functions of $\eta=m^2_{\dS}/(3H^2)$; the top axis gives the mass $m^2_{\AdS}L^2=-3\eta$ of the partner mode. In the window $0\le\eta\le\frac34$ the weights are real, the dS mode lies in the complementary series and its AdS partner obeys the BF bound $m^2_{\AdS}L^2\ge-\frac94$. For $\eta>\frac34$ the weights are complex with $\text{Re}\,\Delta_\pm=\frac32$, the dS mode lies in the principal series, which can produce oscillatory squeezed-limit signals, and the AdS partner violates the BF bound. This is the region of heavy moduli, including those of KKLT ($\eta>124$ for the illustration of section~\ref{sec:kklt}). For $\eta<0$ the dS member is tachyonic. The dots mark the explicit pair of section~\ref{sec:example}; the bracket marks the range $|\eta_V|\ll1$ of a slow-roll inflaton for orientation only, since the vacuum window does not apply directly to a rolling background.}
\label{fig:window}
\end{figure}

\subsection{The moduli mass dichotomy}

The stability window separates string-derived dS vacua into two classes. Combining Eq.~(\ref{eq:window}) with the domain condition $H<1/\LB$, every scalar of a pair with two stable vacua satisfies
\begin{equation}\label{eq:criterion}
m^2\ \le\ \frac94H^2\ <\ \frac{9}{4\LB^2}\ \le\ \frac94\,m_{3/2}^2 .
\end{equation}
The middle bound involves only the background functions $(k,w)$ evaluated at the vacuum; the uplift enters through the position of the vacuum and through the masses. A stable partner therefore requires every scalar to be lighter than $3/(2\LB)$, the scale set by the curvature of the background, and a fortiori lighter than $\frac32m_{3/2}$.

\begin{table}[tb]
\centering\small
\renewcommand{\arraystretch}{1.3}
\begin{tabular}{@{}lll>{\raggedright\arraybackslash}p{3.9cm}@{}}
\toprule
class & condition at the dS vacuum & AdS partner & examples\\
\midrule
no partner & $H\ge1/\LB$ & no positive image & KKLT beyond the edge $u=2$\\
heavy & $H<1/\LB$, some $m>\frac32H$ & ghost-free, BF-unstable & KKLT; large volume with its moduli retained; late-time vacua\\
light & $H<1/\LB$, all $0\le m^2\le\frac94H^2$ & ghost-free, BF-stable & sections~\ref{sec:example} and~\ref{sec:multivacuum}\\
\bottomrule
\end{tabular}
\caption{Classification of tachyon-free dS critical points with $\mathcal{B}<0$ and a goldstino (anti-brane) uplift by their AdS partners under the fixed-background reflection $\RR$, which keeps $k$, $w$ and $f$. For $\mathcal{B}\ge0$ there is no positive image, and tachyonic dS critical points are paired mode by mode as in section~\ref{sec:vacua}. The data required are the background energy $\mathcal{B}=-3/\LB^2$, the Hubble rate and the scalar spectrum at the vacuum. A necessary condition for the light class is Eq.~(\ref{eq:criterion}).}
\label{tab:classes}
\end{table}

If at least one scalar is heavier than $\frac32H$, the partner, when it exists, is ghost-free but violates the BF bound. This is the generic situation for moduli stabilized well above the Hubble scale, and in particular for vacua with a Hubble rate as small as the present one. If all scalars obey Eq.~(\ref{eq:window}), both vacua pass the linear scalar stability test. Figure~\ref{fig:window} shows the two classes on the axis of $\eta$: the light class is the window, the heavy class its complement at $\eta>\frac34$. The light class thus requires a spectrum in which no scalar exceeds $H$ by more than a factor $\frac32$, the regime of quasi-single-field dynamics \cite{Chen:2009we,Chen:2009zp}. Heavy fields can be removed from the four-dimensional theory by constrained superfields \cite{Ferrara:2015tyn,Carrasco:2015iij}, and the window then applies to the fields that remain. A constrained theory is not equivalent to integrating out a stable heavy mode in both members, however: the reflection turns a heavy dS direction into an AdS direction with $m^2L^2\ll-\frac94$, and deleting that direction does not establish the stability of the reflected theory in which it is retained. The resulting classes are summarized in table~\ref{tab:classes} and discussed in section~\ref{sec:classification}.

\subsection{KKLT in closed form}\label{sec:kklt}

Consider the KKLT model \cite{Kachru:2003aw} with a nilpotent uplift \cite{Ferrara:2014kva,Kallosh:2014wsa},
\begin{equation}\label{eq:KKLT}
K=-3\ln\big(T+\bar T-\Xh\Xhb\big),\qquad W=W_0+Ae^{-aT}+\mu^2\Xh,
\end{equation}
with real constants $W_0<0$, $A>0$, $a>0$ and $\mu^2>0$, and $T=\sigma+i\theta$. In the notation of section~\ref{sec:construction}, $k=-3\ln(2\sigma)$, $h=3/(2\sigma)$, $w=W_0+Ae^{-aT}$ and $f=\mu^2$, so that the uplift energy $U=\mu^4/(12\sigma^2)$ does not depend on the axion $\theta$. The background energy is the KKLT potential without uplift,
\begin{equation}\label{eq:kkltB}
\mathcal{B}(\sigma,\theta)=\frac{aAe^{-a\sigma}}{2\sigma^2}\Big[Ae^{-a\sigma}\Big(1+\frac{a\sigma}{3}\Big)+W_0\cos(a\theta)\Big].
\end{equation}
At a critical point with $\theta=0$ all quantities can be expressed in closed form through the two variables
\begin{equation}\label{eq:xb}
x\equiv a\sigma,\qquad b\equiv-\frac{W_0\,e^{a\sigma}}{A},
\end{equation}
the instanton action and the ratio of the flux superpotential to the non-perturbative term. The background energy is negative for $b>1+x/3$. The condition $\partial_\sigma V=0$ fixes the uplift energy, $U=\frac12\sigma\,\partial_\sigma\mathcal{B}$, and the uplift fraction $u=U/U_*$ of section~\ref{sec:reflection} is
\begin{equation}\label{eq:kkltu}
u=\frac{(x+2)(3b-2x-3)}{2(3b-x-3)} .
\end{equation}
It vanishes at the supersymmetric AdS vacuum, $b=1+\frac23x$, where $D_TW=0$, and increases monotonically with $b$ at fixed $x$. It reaches the Minkowski value $u=1$ at $b=\frac13(2x+5)$ and the edge $u=2$ of the domain of $\RR$ at
\begin{equation}\label{eq:bedge}
b_{\rm edge}=\frac{2x^2+3x-6}{3(x-2)} .
\end{equation}
For $x>2$ the dS critical points with a ghost-free partner are therefore those with $\frac13(2x+5)<b<b_{\rm edge}$; they are local minima where in addition $b<b_{\rm m}$, Eq.~(\ref{eq:kkltmerge}) below. With the canonical normalization fixed by $K_{T\bar T}=3/(4\sigma^2)$, the squared masses of $\sigma$ and $\theta$ and the gravitino mass in units of $\LB$ are
\begin{align}
m_\sigma^2\LB^2&=\frac{2x\big[4x^2+10x+5-3b(x+1)\big]}{3b-x-3},\label{eq:kkltsigma}\\
m_\theta^2\LB^2&=\frac{2x^2\,b}{b-1-x/3},\label{eq:kkltaxion}\\
m_{3/2}^2\LB^2&=\frac{3(b-1)^2}{4x\,(b-1-x/3)}\ \ge\ 1.\label{eq:kkltgravitino}
\end{align}
At fixed $x$ both scalar masses decrease monotonically with $b$, so that for $x>2$ and over the whole uplift domain $1+\frac23x<b<b_{\rm edge}$ they are bounded by their values at the two ends,
\begin{align}
\frac{2(2x^3-3x^2-12x-4)}{x+2}&<m_\sigma^2\LB^2\le2(x+2)(2x+1),\nonumber\\
\frac{2x(2x^2+3x-6)}{x+2}&<m_\theta^2\LB^2\le2x(2x+3).\label{eq:kkltbounds}
\end{align}
Both moduli therefore have, for every uplift for which a partner exists,
\begin{equation}\label{eq:kkltasym}
m_\sigma^2\LB^2=4(a\sigma)^2\big[1+\mathcal{O}(1/a\sigma)\big],\qquad m_\theta^2\LB^2=4(a\sigma)^2\big[1+\mathcal{O}(1/a\sigma)\big],
\end{equation}
far above the value $\frac94$ allowed by the criterion (\ref{eq:criterion}). In addition, Eq.~(\ref{eq:kkltaxion}) gives the exact bound $m_\theta^2\LB^2>2(a\sigma)^2$ at every critical point with negative background energy, whatever the uplift. For $a\sigma>3/(2\sqrt2)$ the criterion is therefore violated at every critical point in the domain of $\RR$, and controlled one-exponential KKLT has $a\sigma\gg1$. Since the dS member has $H^2\LB^2=u-1<1$, both moduli lie even further outside the window (\ref{eq:window}) in terms of $\eta$.

The closed forms make the origin of this behavior transparent. At the supersymmetric point $m_{3/2}\LB=1$ and the moduli masses are $m_\sigma\simeq m_\theta\simeq2a\sigma\,m_{3/2}$, the familiar KKLT hierarchy; the uplift lowers the masses in units of $\LB$ by relative corrections of order $1/a\sigma$ only, and at the edge of the domain $m_{3/2}^2\LB^2=x^2/(x^2-4)$ remains close to one. Since $a\sigma$ grows logarithmically as $|W_0|\to0$, the violation becomes stronger in the regime of parametric control. The dS minimum disappears where $m_\sigma^2$ vanishes, at
\begin{equation}\label{eq:kkltmerge}
b_{\rm m}=\frac{4x^2+10x+5}{3(x+1)},\qquad u_{\rm m}=\frac{(x+2)^2(2x+1)}{2(3x^2+6x+2)}=\frac x3+\frac56+\mathcal{O}(1/x).
\end{equation}
Since $b_{\rm m}$ exceeds $b_{\rm edge}$ precisely for $x>x_c\simeq3.42$, all dS critical points in the domain of $\RR$ are local minima there, and in particular in the controlled regime $x\gg1$. Between the edge $u=2$ and $u_{\rm m}$ the dS minimum then persists, but $\RR(h_+)$ is negative at the vacuum and the fixed-background map has no positive image.

As an illustration, take $W_0=-10^{-4}$, $A=1$ and $a=0.1$. The supersymmetric vacuum lies at $\sigma\simeq113.6$, the Minkowski vacuum at $\sigma\simeq114.4$ and the edge of the domain at $\sigma\simeq115.4$, so that all dS vacua with a ghost-free partner lie in the narrow interval $114.4<\sigma<115.4$. Along this branch $m_\sigma^2\LB^2$ decreases from $504$ to $374$ and $m_\theta^2\LB^2$ from $543$ to $503$, to be compared with $\frac94$; correspondingly the two moduli have $\eta_\sigma>124$ and $\eta_\theta>167$ in the notation of Eq.~(\ref{eq:window}), while $m_{3/2}^2/H^2$ decreases from infinity at the Minkowski end to $1.03$ at the edge. Every dS vacuum of the model is therefore either without a ghost-free partner or paired with a BF-unstable AdS vacuum. The argument uses this K\"ahler potential, superpotential and uplift; racetrack superpotentials, moduli mixing, axion-dependent uplifts and corrections to the K\"ahler potential require their own analysis.

\subsection{Large volume and light spectra}

In the large-volume scenario \cite{Balasubramanian:2005zx,Conlon:2005ki} the background is a non-supersymmetric AdS minimum at volume $\mathcal{V}\gg1$, with $|\mathcal{B}|\sim m_{3/2}^2/\mathcal{V}$ up to logarithmic factors. The volume modulus has a mass of order $m_{3/2}/\sqrt{\mathcal{V}}$ \cite{Conlon:2005ki}, which is of order $1/\LB$, so that its position relative to the window depends on numerical factors. The small K\"ahler modulus, the complex-structure moduli and the dilaton, however, have masses of order $m_{3/2}$ or above \cite{Conlon:2005ki}, larger than $3/(2\LB)$ by a power of $\mathcal{V}$. In an uplift that retains this hierarchy, large-volume dS vacua therefore also belong to the heavy class; other spectra and constrained theories require a separate analysis. The light class is realized by the explicit pair of section~\ref{sec:example}, where both scalars have $\eta=\frac12$, and by the final vacua of the inflationary pair of section~\ref{sec:multivacuum}. Rolling backgrounds, to which a vacuum BF test does not directly apply, are discussed in section~\ref{sec:inflation}.

\subsection{Classification and swampland criteria}\label{sec:classification}

These results provide a classification of dS vacua of string-derived effective theories with an anti-brane uplift, summarized in table~\ref{tab:classes}. It requires only the background energy, the Hubble rate and the scalar spectrum at the vacuum. If $H\ge1/\LB$ the fixed-background map has no positive image. If $H<1/\LB$ and some scalar is heavier than $\frac32H$, the partner is ghost-free and BF-unstable. If $H<1/\LB$ and all scalars obey Eq.~(\ref{eq:window}), the partner is a non-supersymmetric AdS vacuum that passes the linear scalar stability test.

The correspondence also isolates what swampland criteria can constrain. In its refined form the de Sitter conjecture \cite{Obied:2018sgi,Ooguri:2018wrx} requires, wherever $V>0$, either $|\nabla V|\ge c\,V$ or $\min\nabla\nabla V\le-c'V$, with positive constants $c$ and $c'$ of order one, where $\min\nabla\nabla V$ is the smallest eigenvalue of the covariant Hessian; for $V<0$ the first condition holds trivially. Within the pairs, the two members share the visible kinetic metric, the superpotential, the surviving fermion operators and $|V|$ with all its $|V|$-normalized derivatives, and differ only in the sign of the potential, which is set by the goldstino metric. The de Sitter conjecture, trivially satisfied by the AdS member, therefore cannot be traced to any shared data: within this class it constrains the goldstino (anti-brane) sector alone, that is, whether the over-uplifted metric $h_+$ admits an ultraviolet completion. Since the two members are distinct effective theories related by a classical correspondence, the AdS member does not lift the constraints of the conjecture on the dS member: the correspondence relocates the question rather than resolving it.

\begin{table}[t]
\centering\small
\renewcommand{\arraystretch}{1.3}
\begin{tabular}{@{}>{\raggedright\arraybackslash}p{3.15cm}>{\raggedright\arraybackslash}p{3.6cm}>{\raggedright\arraybackslash}p{3.0cm}>{\raggedright\arraybackslash}p{3.6cm}@{}}
\toprule
criterion & dS member, $V_+=P$ & AdS member, $V_-=-P$ & image under $\RR$\\
\midrule
slope branch\newline $|\nabla V|\ge c\,V$ & constraining; at $N=60$ it requires $c\le\sqrt{2\epsilon_V}\simeq0.008$ & holds trivially & $|\nabla V|/|V|$ is common to both members\\
\addlinespace
Hessian branch\newline $\min\nabla\nabla V\le-c'V$ & at a critical point it requires $\eta\le-c'$; violated at every dS minimum & not needed; on the image of the plateau it reads $-\eta_V\le c'$ \cite{Atli:2020dni} & $\eta=-m^2_{\AdS}L^2/3$; it requires an AdS direction with $\Delta_+(\Delta_+-3)\ge3c'$\\
\addlinespace
non-supersymmetric\newline AdS vacua \cite{Ooguri:2016pdq} & --- & applies: $F^{\Xh}\neq0$ at every critical point & BF-stable partners must decay non-perturbatively\\
\bottomrule
\end{tabular}
\caption{Swampland criteria across the correspondence. The de Sitter conjecture \cite{Obied:2018sgi,Ooguri:2018wrx} applies where $V>0$ and requires one of its two branches, with constants $c,c'$ of order one; $\eta=m^2_{\dS}/(3H^2)$ at critical points and $\eta_V$ along the inflationary slice. The two members share $|V|$ and its derivatives and differ only in the sign of the potential, which is set by the goldstino metric, so that the criteria constrain the goldstino sector. The number refers to the inflationary pair of section~\ref{sec:multivacuum}; the entry for non-supersymmetric AdS vacua assumes the conjecture of Ref.~\cite{Ooguri:2016pdq}.}
\label{tab:swampland}
\end{table}

Table~\ref{tab:swampland} shows how the criteria act on the two members. The ratio $|\nabla V|/|V|$ is common to both, so that the slope branch constrains the dS member exactly as it would any theory with potential $P$. At a dS critical point the slope branch fails, and the Hessian branch requires a tachyonic direction with $\eta\le-c'$. Through Eq.~(\ref{eq:masses}) this is the statement that the AdS member has a direction with $m^2_{\AdS}L^2=-3\eta\ge3c'>0$, so that its critical point is not a maximum in all directions. A metastable dS member conflicts with the conjecture, as does any metastable dS vacuum of a theory to which the conjecture is meant to apply. For the inflationary pair of section~\ref{sec:multivacuum} the dS member requires $c\le\sqrt{2\epsilon_V}\simeq0.008$ or $c'\le|\eta_V|\simeq0.014$ at $N=60$ e-folds, and its final dS minimum, with $\nabla V=0$ and $\eta=\frac{8}{45}$, violates both branches. These numbers are fixed by the data rather than by the benchmark: to first order in slow roll the conjecture holds at the pivot scale only if $c\le\sqrt{r/8}$ or $c'\le\frac12(1-n_s)-\frac{3}{16}\,r$, that is $c\lesssim0.07$ or $c'\lesssim0.02$ for any single-field model compatible with the CMB data \cite{Planck:2018jri,BICEP:2021xfz}, in line with Refs.~\cite{Agrawal:2018own,Kinney:2018nny}, and on the inverse-quartic plateau of the pair $-\eta_V\simeq5/(6N)$ for every $\kappa$, $v$ and $\Lambda/V_0$. The violation at the final vacuum holds for every $\Lambda>0$, which is also the condition for the domain wall of the AdS member in section~\ref{sec:multivacuum} to have an ultraviolet end at a non-supersymmetric AdS critical point, so that the de Sitter and non-supersymmetric AdS criteria of table~\ref{tab:swampland} act on the same point of the pair. Conversely, non-supersymmetric AdS vacua are conjectured to be unstable \cite{Ooguri:2016pdq}. The AdS member is never supersymmetric at finite $h_-$. If the conjecture holds, a pair in the light class, whose AdS member passes the linear test, would therefore require a non-perturbative instability of that member in any ultraviolet completion, whereas pairs in the heavy class are unstable already at the perturbative level.

This also gives the motivation of Ref.~\cite{Atli:2020dni} a precise place. There a superconformal sign choice produced the negative potential $-f^2$ from a K\"ahler function with negative-definite metric, and the Hessian branch evaluated on it, together with a stability condition, gave the range $0\lesssim-\eta_V\le c'$. By section~\ref{sec:signs} and table~\ref{tab:signs}, that potential belongs to a ghost-ridden relabeling of the inflationary theory. At $W\neq0$ the same range is realized without ghosts. Along the image of the inflationary trajectory the healthy AdS member $h_-=\RR(h_+)$ has $V_-=-V_+$ and inflaton Hessian $\partial_\varphi^2V_-=-\eta_V|V_-|$, so that in the inflaton direction the Hessian branch, $\min\nabla\nabla V_-\le-c'V_-=c'|V_-|$, reads $-\eta_V\le c'$, and on a concave plateau it becomes the range $0\lesssim-\eta_V\le c'$ of Ref.~\cite{Atli:2020dni}. As for any AdS theory, the conjecture is satisfied there already through its first branch, and it continues to constrain the dS member on its own.

\section{Inflationary Cosmologies and Their AdS Domain-Wall Partners}\label{sec:inflation}

Many supergravity models of inflation contain an inflaton multiplet $\Phi$ and a stabilizer multiplet $S$ with $W=Sf(\Phi)$, so that $W=0$ along the inflationary trajectory $S=0$ \cite{Kallosh:2010xz}. By Eq.~(\ref{eq:W0pos}) and appendix~\ref{app:W0} they have no healthy partner with the opposite potential that also remains on the $W=0$ locus; their stabilizer spectra, including the healthy branch of the model of Ref.~\cite{Atli:2020dni}, are collected in appendix~\ref{app:stabilizers}. Their nilpotent versions have $w\neq0$, and $\RR$ acts on them. Throughout this section $f$ is a real holomorphic function, $f(\bar\Phi)=\overline{f(\Phi)}$, up to a removable constant phase, so that the real slice $\text{Im}\,\Phi=0$ is a consistent truncation. On rolling configurations we write $V=3H_V^2$; the Hubble rate $H$ differs from $H_V$ by the kinetic energy of the canonically normalized inflaton $\varphi$, $3H^2=V+\frac12\dot\varphi^2$. The slow-roll parameters are
\begin{equation}\label{eq:slowroll}
\epsilon_V\equiv\frac12\Big(\frac{\partial_\varphi V}{V}\Big)^2,\qquad \eta_V\equiv\frac{\partial_\varphi^2V}{V}.
\end{equation}
Masses below are covariant Hessians of the potential in canonical normalization, which at critical points are the physical masses; the fluctuation operator of a rolling background contains further gravitational and velocity-dependent terms.

\subsection{Nilpotent versions and their AdS domain-wall partners}

Replacing the stabilizer by the goldstino multiplet removes the scalar partner of the goldstino \cite{Ferrara:2014kva,Kallosh:2014via}, and a non-vanishing $w(\Phi)$ brings the potential to the form (\ref{eq:VAB}), on which $\RR$ acts. In the simplest version \cite{Ferrara:2014kva},
\begin{equation}\label{eq:nil1}
K=-\tfrac12(\Phi-\bar\Phi)^2+\Xh\Xhb,\qquad W=W_0+\Xh f(\Phi),
\end{equation}
with a real constant $W_0$, one has $D_\Phi w=0$ along the real axis, where $\mathcal{B}=-3W_0^2$, $\mathcal{A}=f^2$ and the inflationary potential is $V_+=f^2-3W_0^2$; the constant $W_0$ tunes the vacuum energy after inflation \cite{Ferrara:2014kva}. On the real axis the AdS partner has
\begin{equation}\label{eq:nil2}
h_-=\RR(1)=\frac{f^2}{6W_0^2-f^2},\qquad V_-=3W_0^2-f^2,
\end{equation}
and its metric is positive along the trajectory precisely when $f^2<6W_0^2$, that is $H_V<m_{3/2}=|W_0|$ wherever $V_+>0$, in agreement with Eq.~(\ref{eq:m32dS}); a strict inequality on a compact segment extends positivity to a neighborhood of the trajectory. By Eq.~(\ref{eq:LcosLwall}) the inflationary trajectory of the dS member is a flat domain wall of the AdS member, and the covariant Hessians of the two members are opposite at common field points. For the second scalar of $\Phi$ the dS member has
\begin{equation}\label{eq:nil3}
m_\beta^2=3H_V^2\Big[2-\eta_V+\frac{2\epsilon_VH_V^2}{H_V^2+m_{3/2}^2}\Big]+4m_{3/2}^2 .
\end{equation}
Throughout slow roll ($|\eta_V|\ll1$) inside the domain of $\RR$, where $m_{3/2}>H_V$, the bracket is positive and the last term alone exceeds $\frac94H_V^2$, so that $m_\beta^2>\frac94H_V^2$. In this regime $\beta$ is a heavy transverse direction, the inflationary counterpart of the heavy class of section~\ref{sec:strings}: at an exact paired critical point with these masses the partner would violate the BF bound. Along a rolling wall, however, there is no pointwise BF theorem obtained by replacing $L^{-2}$ with $|V|/3$; the stability of the wall is decided by its gauge-invariant radial fluctuation problem with boundary conditions, which slow roll alone does not settle.

Orthogonal nilpotent superfields, $\Xh^2=0$ and $\Xh(\Phi-\bar\Phi)=0$ \cite{Ferrara:2015tyn,Carrasco:2015iij}, define a different constrained theory in which $\beta$ and its fermionic partner are absent. They also remove the auxiliary contribution of $\Phi$, so that the potential is $e^k(|f|^2/h-3|w|^2)$ and the background energy is $\mathcal{B}_{\rm orth}=-3e^k|w|^2$ \cite{Ferrara:2015tyn}. For the model (\ref{eq:nil1}) this agrees with $V_+$ on the real trajectory, and $\RR$ applies with $\mathcal{B}_{\rm orth}$. The inflaton is then the only scalar, and its reflected Hessian $-3\eta_VH_V^2$ satisfies the BF inequality in units of the local curvature whenever $\eta_V\le\frac34$, which holds throughout slow roll; as for any rolling background, this is a statement about the potential rather than a stability theorem for the wall. Alternatively, a dependence of the goldstino metric on $\Phi-\bar\Phi$ can bring $\beta$ into the complementary series, where it leaves a power-law imprint (\ref{eq:squeezed}) on the squeezed limit. Near the end of inflation $\eta_V$ grows. Since $\RR$ fixes the zeros of $P$ but reverses its curvature, the post-inflationary Minkowski minimum of the dS member is a Minkowski maximum of its partner.

More generally, for any inflationary profile $P(\Phi,\bar\Phi)$ with $P<-\mathcal{B}$, Eq.~(\ref{eq:hpm}) produces an inflationary dS member and a ghost-free AdS member with $V=-P$, given by the Lagrangians (\ref{eq:L_dS})--(\ref{eq:L_AdS}). Its transverse Hessian is the negative of that of the dS member, and at any exact critical point the window (\ref{eq:window}) decides whether both members are stable.

\subsection{A closed-form inflationary pair}\label{sec:multivacuum}

The pairings of section~\ref{sec:vacua} can all be realized within one closed-form pair, which also supports slow-roll inflation with a graceful exit. Take a single inflaton chiral multiplet $\Phi=(\varphi+i\beta)/\sqrt2$ with the shift-symmetric K\"ahler potential, a constant superpotential and the goldstino metrics $h_\pm$,
\begin{equation}\label{eq:mv_model}
k=-\tfrac12(\Phi-\bar\Phi)^2=\beta^2,\qquad W=M+F\Xh,\qquad K_\pm=k+h_\pm\Xh\Xhb,
\end{equation}
with real constants $M,F\neq0$. One finds $\mathcal{A}=e^{\beta^2}F^2$ and $\mathcal{B}=e^{\beta^2}M^2(2\beta^2-3)$, and for any target profile $P$ the metrics $h_\pm=\mathcal{A}/(\pm P-\mathcal{B})$ of Eq.~(\ref{eq:hpm}) give $V_\pm=\pm P$ exactly; the inflaton potential is carried by the goldstino metric, as in Ref.~\cite{Kallosh:2017wnt}. The kinetic metric $k_{\Phi\bar\Phi}=1$ is flat, so that $\varphi$ and $\beta$ are canonically normalized and covariant Hessians are ordinary second derivatives. We take the profile
\begin{equation}\label{eq:mv_profile}
P(\varphi,\beta)=P(\varphi)+\frac12\,\beta^2\,\partial_\varphi^2P(\varphi),\qquad P(\varphi)=\Lambda+V_0\,\frac{s^2(1+\kappa s^2)}{1+\kappa s^4},\qquad s\equiv1-\frac{\varphi^2}{v^2},
\end{equation}
with constants $V_0>0$, $v>0$, $\Lambda\ge0$ and $\kappa\ge0$. It is even in $\beta$, so that the real slice $\beta=0$ is a consistent truncation, along which $V_\pm=\pm P(\varphi)$ and $h_\pm=F^2/[3M^2\pm P(\varphi)]$, and $\partial_\beta^2P=\partial_\varphi^2P$ on the slice, so that at its critical points both scalars have the same mass. Since $\Lambda+V_0-P(\varphi)=V_0(1-s^2)/(1+\kappa s^4)$, one has $\Lambda\le P(\varphi)\le\Lambda+V_0$ for $|\varphi|\le\sqrt2\,v$, and $P$ decreases monotonically from $\varphi=0$ to $|\varphi|=v$. For $3M^2>\Lambda+V_0$ both metrics are therefore positive on this interval and, by continuity, in a tube around it; the gravitino mass $m_{3/2}=|M|$ then exceeds $H_V$ on the slice, as required by Eq.~(\ref{eq:m32dS}).

\begin{figure}[t]
\centering
\includegraphics{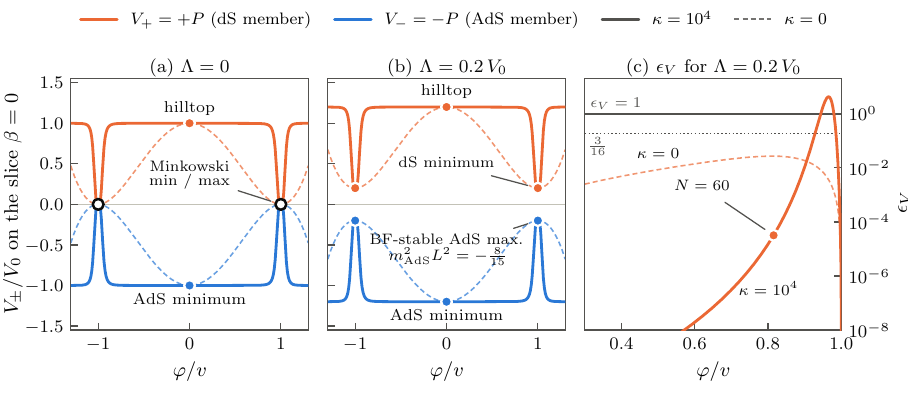}
\caption{The inflationary pair (\ref{eq:mv_model})--(\ref{eq:mv_profile}) along the slice $\beta=0$ for $v=15$ and $\kappa=10^4$ (solid), with the quartic $\kappa=0$ (dashed) for comparison; at all critical points the transverse field $\beta$ has the same mass as $\varphi$. (a) $\Lambda=0$: the dS member $V_+=+P$ has a plateau with a hilltop at $\varphi=0$ and Minkowski minima at $\varphi=\pm v$; the AdS member $V_-=-P$ has a genuine minimum at $\varphi=0$ and Minkowski maxima at $\varphi=\pm v$, where the two potentials touch tangentially. (b) $\Lambda=0.2\,V_0$: the endpoints are dS minima of the dS member and AdS maxima of the AdS member with $m^2_{\AdS}L^2=-\frac{8}{15}>-\frac94$ for every $\kappa$, linearly stable in both scalar directions. (c) Slow-roll parameter $\epsilon_V$ of the dS member for $\Lambda=0.2\,V_0$: the quartic stays below the bound $\frac3{16}$ implied by the window, whereas for $\kappa=10^4$ the steep transition ends inflation; the dot marks the point $N=60$ e-folds before $\epsilon_V=1$, with the leading slow-roll estimates $n_s\simeq0.972$ and $r\simeq5\times10^{-4}$.}
\label{fig:multivacuum}
\end{figure}

For $\kappa=0$ the profile is the quartic $\Lambda+V_0(1-\varphi^2/v^2)^2$. The parameter $\kappa$ separates two scales. Near the endpoints, with $\delta\equiv1-|\varphi|/v$,
\begin{equation}\label{eq:mv_twoscale}
P(\varphi)=\Lambda+V_0\big[4\delta^2-4\delta^3+(1+16\kappa)\,\delta^4+\mathcal{O}(\delta^5)\big],
\end{equation}
so that the value and the curvature of $P$ at $|\varphi|=v$ do not depend on $\kappa$, while $\kappa$ controls the steepness above them. For $\kappa\gg1$ the profile is a plateau, $P\simeq\Lambda+V_0[1-(1-s^2)/(\kappa s^4)]$, which drops to $\Lambda$ in a steep transition at $|s|\lesssim\kappa^{-1/4}$. The only critical points on the segment $|\varphi|\le v$ are the origin and the endpoints, with
\begin{align}
V_\pm(0,0)&=\pm(\Lambda+V_0),&m^2_{\varphi,\pm}&=m^2_{\beta,\pm}=\mp\frac{4V_0}{(1+\kappa)\,v^2},\label{eq:mv_origin}\\
V_\pm(\pm v,0)&=\pm\Lambda,&m^2_{\varphi,\pm}&=m^2_{\beta,\pm}=\pm\frac{8V_0}{v^2}.\label{eq:mv_endpoints}
\end{align}
The dS member has a hilltop at the origin, flattened by the factor $1/(1+\kappa)$, and the AdS member a genuine minimum in both directions. For $\Lambda=0$ the endpoints are Minkowski minima of the dS member and Minkowski maxima of the AdS member, at which $V_\pm=0$ and $h_+=h_-=h_*=F^2/(3M^2)$: the two potentials touch tangentially, as described in section~\ref{sec:reflection}. For $\Lambda>0$ they are dS minima of the dS member, with $V_+=\Lambda=3H^2$ and $\eta=8V_0/(v^2\Lambda)$, and AdS maxima of the AdS member, with $V_-=-\Lambda=-3/L^2$. These are BF-stable in both scalar directions when
\begin{equation}\label{eq:mv_BF}
m^2_{\AdS}L^2=-\frac{24V_0}{v^2\Lambda}\ \ge\ -\frac94\qquad\Longleftrightarrow\qquad \frac{8V_0}{v^2}\le\frac34\Lambda,
\end{equation}
which is the window (\ref{eq:window}) for the dS minima and does not involve $\kappa$. The dS member thus contains a hilltop connected to Minkowski or dS minima, and the AdS member a genuine minimum together with Minkowski maxima or BF-stable AdS maxima; for $\Lambda>0$ the endpoints belong to the light class of section~\ref{sec:strings}. Figure~\ref{fig:multivacuum} shows both cases for $v=15$ and $\kappa=10^4$; for $\Lambda=0.2\,V_0$ one finds $m^2_{\AdS}L^2=-\frac{8}{15}$ at the endpoints and $\eta_V\simeq-1.5\times10^{-6}$ at the hilltop. A single profile with separate Minkowski and dS minima is given in appendix~\ref{app:multivacuum}.

\paragraph{Graceful exit and BF stability.} The BF condition (\ref{eq:mv_BF}) constrains only the curvature of the final minimum, and $\kappa$ decouples it from the slope that ends inflation. With $q\equiv\varphi/v$ and $\lambda\equiv\Lambda/V_0$, the quartic member has
\begin{equation}\label{eq:mv_eps}
\epsilon_V\big|_{\kappa=0}=\frac{8q^2(1-q^2)^2}{v^2\big[\lambda+(1-q^2)^2\big]^2}\ \le\ \frac{2q^2}{v^2\lambda},
\end{equation}
so that for $\Lambda>0$ the BF condition (\ref{eq:mv_BF}) gives $\epsilon_V\le\frac{3}{16}$ on the whole segment, and the accelerating solution approaches the dS minimum without a decelerating stage. This is not an obstruction of the BF condition itself, which constrains only the quadratic basin of the final vacuum: where $V\simeq\Lambda+\frac12m^2\delta\varphi^2$, one has $\epsilon_V\le m^2/(4\Lambda)$, which the window bounds by $\frac{3}{16}$. Inflation can therefore end only above this basin, and the quartic fails because its single scale $v$ sets both the curvature of the minimum and the slope above it. For $\kappa\gg1$ the basin is narrow, $\delta\lesssim1/(2\sqrt\kappa)$, and the transition above it is steep. For $v=15$ and $\lambda=0.2$ the slow-roll parameter $\epsilon_V$ exceeds one for $\kappa\gtrsim650$. For $\kappa=10^4$ it rises from values of order $10^{-5}$ on the plateau to one at $|\varphi|\simeq0.944\,v$, as shown in figure~\ref{fig:multivacuum}(c).

\paragraph{Spectral predictions.} On the plateau $P$ approaches $\Lambda+V_0$ as an inverse fourth power of $s$, like the D3--anti-D3 Coulomb term of brane inflation \cite{Kachru:2003sx}, so that $n_s\simeq1-5/(3N)$. At leading order in slow roll, the adiabatic mode on the straight trajectory $\beta=0$ has $n_s\simeq0.972$ and $r\simeq5\times10^{-4}$ at $N=60$ e-folds before $\epsilon_V=1$, where $\varphi\simeq12.230$, and $n_s\simeq0.967$ at $N=50$. Counting $N$ from the first crossing $\epsilon_H=1$ of the full background instead, a numerical solution of the Mukhanov--Sasaki equation gives $(n_s,r)\simeq(0.9721,\,5.2\times10^{-4})$ at $N=60$ and $(0.9665,\,7.1\times10^{-4})$ at $N=50$. These values are compatible within $2\sigma$ with the Planck 2018 and BICEP/Keck constraints \cite{Planck:2018jri,BICEP:2021xfz}.

A numerical integration of the background equations from the exact $N=60$ slow-roll point, $\varphi(0)\simeq12.230$ with $d\varphi/dN=-P_{,\varphi}/P$, confirms the exit. The first crossing $\epsilon_H\equiv-\dot H/H^2=1$ occurs after $N\simeq61.62$ e-folds, $\epsilon_H$ reaches a maximum $\epsilon_H\simeq1.42$, and the decelerating stage $\epsilon_H>1$ lasts $\Delta N\simeq0.31$ e-folds. The inflaton then overshoots the minimum once, to $\varphi\simeq1.017\,v$, recrosses it and approaches it monotonically, while the expansion approaches the final dS phase. Since $\Lambda=0.2\,V_0$ is chosen for visibility, the benchmark demonstrates a transient graceful exit from inflation into a final vacuum with a BF-stable reflected partner; a long hot radiation era and a complete thermal history require in addition a scale separation $\Lambda\ll V_0$ and an explicit reheating sector.

The final approach is controlled by the window. By section~\ref{sec:window}, the window $m\le\frac32H_{\rm late}$ at the final minimum, with $H_{\rm late}=(\Lambda/3)^{1/2}$, is the condition that the asymptotic linear relaxation in its quadratic basin is overdamped, with real weights; in the example $m/H_{\rm late}\simeq0.73$, and the late-time approach $v-\varphi\propto e^{-\Delta_-H_{\rm late}t}$ has $\Delta_-\simeq0.19$. The window does not constrain transient anharmonic oscillations outside the quadratic basin. The single overshoot above is an example, and in a steeper well, such as $V\simeq\Lambda+\frac12m^2\delta\varphi^2+\frac14g\,\delta\varphi^4$ with $m\le\frac32H_{\rm late}$ and a dominant quartic term, the field can oscillate many times before it enters the overdamped regime. What a BF-stable partner excludes is sustained harmonic oscillation about the final minimum: final vacua with $m>\frac32H_{\rm late}$, and in particular an inflaton that reheats by oscillating about a late-time minimum with $m\gg H_{\rm late}$, have BF-unstable reflected partners, the inflationary counterpart of the moduli mass dichotomy. With a BF-stable partner, reheating must therefore proceed during the transient stage or through non-oscillatory channels, such as gravitational particle production \cite{Ford:1986sy} or instant preheating \cite{Felder:1998vq}, as in quintessential inflation \cite{Peebles:1998qn} and related models \cite{Felder:1999pv}. In the family (\ref{eq:mv_profile}) the curvature of the minimum is tied to $v$, and for $\Lambda\ll V_0$ the window requires $v^2\ge32V_0/(3\Lambda)$; a final vacuum energy far below the inflationary scale requires a separate scale for the minimum and an inflaton that remains lighter than $\frac32H_{\rm late}$ there.

The real slice is a consistent truncation but not an attractor near the hilltop: the transverse mass $m_\beta^2=\partial_\varphi^2P$ is negative wherever the inflaton potential is concave, which is what makes the origin a genuine minimum of the partner. On the plateau it is small, $|m_\beta^2|=3|\eta_V|H_V^2\ll H_V^2$. Since the turn rate of the straight trajectory $\beta=0$ in the flat field space vanishes, the entropy mode does not source the curvature perturbation at linear order; during the steep transition, however, $m_\beta^2$ drops to about $-8H^2$ and the transverse mode can undergo tachyonic growth, which motivates a full two-field perturbation analysis, including isocurvature modes and initial conditions.

\paragraph{Domain-wall partner and the c-theorem.} By Eq.~(\ref{eq:LcosLwall}) the finite-time inflationary trajectory above, including its graceful exit, is mapped to a segment of a flat domain wall of the AdS member. A complete heteroclinic wall between the two AdS critical points, the minimum at $\varphi=0$ and a BF-stable maximum at $\varphi=\pm v$, which in a holographic setting has the form of a renormalization-group flow from $L^2=3/\Lambda$ to $L^2=3/(\Lambda+V_0)$, is the image of a flat cosmology that leaves the hilltop in the infinite past and approaches a dS minimum, and it is selected by its asymptotic boundary conditions at the two critical points. The dictionary (\ref{eq:Delta}) and Eq.~(\ref{eq:LcosLwall}) give this flow a precise kinematic content, without any assumption about a dual description: the warp factor grows from the AdS minimum, the image of the hilltop, to the BF-stable maximum, the image of the final dS vacuum, which are therefore its infrared and ultraviolet ends. At the infrared end $m^2_{\AdS}L^2=-3\eta_V$ gives $\Delta_+-3\simeq|\eta_V|\simeq1.5\times10^{-6}$, so that slow roll near the hilltop is the image of a nearly marginal irrelevant direction, on which the Hessian branch of the refined de Sitter conjecture would impose $\Delta_+(\Delta_+-3)\ge3c'$ (section~\ref{sec:classification}); at the ultraviolet end $\Delta_+\simeq2.81$ and $\Delta_-\simeq0.19$, the late-time relaxation rate found above.

Between the two ends the effective radius $1/\rho'$ grows monotonically from $L_{\rm IR}$ to $L_{\rm UV}$, with $L_{\rm IR}^2=3/(\Lambda+V_0)$ and $L_{\rm UV}^2=3/\Lambda$, since in the gauge $N=1$ the wall obeys $\rho''=-k_{a\bar b}z'^a\bar z'^{\bar b}\le0$, the image under Eq.~(\ref{eq:LcosLwall}) of $\dot H\le0$. This is the classical holographic c-theorem \cite{Freedman:1999gp}, which requires $k_{a\bar b}>0$ relative to the Einstein--Hilbert term and fails in the relabelings of table~\ref{tab:signs} that produce $-f^2$, and it reproduces the assignment of the inflationary past to an infrared and of the late de Sitter phase to an ultraviolet fixed point proposed in dS/CFT \cite{Strominger:2001gp}. Table~\ref{tab:swampland} summarizes the swampland criteria for this pair.

\section{Conclusions}\label{sec:conclusions}
For two decades, the domain-wall/cosmology correspondence has appeared to exact an unavoidable price in supergravity: the cosmological partner of a supersymmetric domain wall is a non-unitary pseudo-supergravity with vector ghosts \cite{Skenderis:2007sm}. In this paper, we have shown that these ghosts are tied to (pseudo-)supersymmetric cosmologies rather than to supergravity itself. Once supersymmetry is broken spontaneously, the full pointwise $V\leftrightarrow-V$ correspondence is realized between two ordinary, ghost-free four-dimensional $\mathcal{N}=1$ supergravities, and de Sitter and anti-de Sitter potentials need not be separated by ghosts. The correspondence emerges naturally from a two-layer mechanism inside a single superconformal K\"ahler cone. While the global signs of the superconformal action merely relabel the K\"ahler potential to select a ghost-free base geometry, leaving the potential locked to non-negative values on $W=0$ loci, the weight $h$ of a nilpotent goldstino direction provides an independent second control. At $W\neq0$ it balances the negative compensator direction, and the exact $\mathbb{Z}_2$ involution $\RR(h)$ reflects the goldstino metric through the flat theory $h_*$ with identically vanishing potential. The outcome is a pair of explicit Lagrangians, Eqs.~(\ref{eq:L_dS})--(\ref{eq:L_AdS}), sharing the same field content, the same positive visible K\"ahler metric $k_{a\bar b}>0$, positive goldstino metrics $h_\pm>0$, and identical surviving unitary-gauge ($\chi^{\Xh}=0$) fermion operators while carrying opposite field-dependent potentials, thereby realizing the domain-wall/cosmology correspondence between two ordinary, ghost-free supergravities.

The paired theories obey a sharp dictionary: $H=1/L$, $m^2_{\dS}=-m^2_{\AdS}$, equal conformal weights, and a common gravitino mass with $m_{3/2}>H$. Its stability window, $0\le m^2\le\frac94H^2$, unifies physical questions that are usually treated separately. In holography, it identifies the Breitenlohner--Freedman (BF) bound of the AdS member with the reality of the dS/CFT conformal weights. In primordial cosmology, it separates the power-law squeezed-limit bispectrum of light fields from the oscillatory signals of the cosmological collider, and it determines whether the asymptotic linear relaxation of a scalar field in the quadratic basin of a dS vacuum is overdamped or oscillatory. In string flux compactifications, it establishes a sharp moduli mass dichotomy: the reflection pairs over-uplifted dS vacua with under-uplifted AdS vacua, and a BF-stable partner requires every scalar to be lighter than $3/(2\LB)$, a scale fixed solely by the background functions at the vacuum. In one-modulus KKLT, our closed-form spectrum places both moduli a factor of order $a\sigma\gg1$ above this threshold across the entire uplift domain, excluding a stable partner, while large-volume vacua whose uplift retains the moduli mass hierarchy fall into the same heavy class. At the same time, our analysis establishes exact selection rules that explain why earlier approaches encountered ghosts: on $W=0$ loci no covariant rearrangement can produce a negative potential without flipping a kinetic sign, a visible K\"ahler function of mixed signature cannot be cured by global superconformal signs, and the AdS partner breaks supersymmetry at every finite goldstino metric, so that a supersymmetric AdS vacuum is never mapped onto a supersymmetric dS vacuum, consistent with the absence of unitary de Sitter superalgebras.

Inflationary cosmology demonstrates the power of the correspondence in action. At the level of critical points, the reflection pairs dS minima inside the stability window with BF-stable AdS maxima, slow-roll hilltops with genuine AdS minima, and Minkowski minima with Minkowski maxima. At the level of full trajectories, flat inflationary cosmologies map directly onto healthy AdS domain walls. Because the BF bound constrains only the curvature of the final vacuum rather than the slope that ends inflation, our closed-form two-scale profile achieves both a realistic slow-roll plateau ($n_s\simeq0.97$, $r\simeq5\times10^{-4}$) and a steep graceful exit into a dS minimum with a BF-stable AdS partner. Furthermore, because the correspondence preserves the positivity of the visible K\"ahler metric $k_{a\bar b}>0$, the partner domain wall automatically obeys the classical holographic $c$-theorem ($\rho''\le0$ as the image of $\dot H\le0$), in sharp contrast to the ghost-ridden $W=0$ relabelings. The stability window also gives the end of inflation a clear physical meaning: a BF-stable partner of the final vacuum requires the asymptotic linear relaxation in its quadratic basin to be overdamped, whereas final vacua around which the inflaton oscillates harmonically ($m>\frac32H_{\rm late}$, and in particular $m\gg H_{\rm late}$) have BF-unstable partners, although transient anharmonic oscillations outside the quadratic basin remain unconstrained. Whether the domain walls paired with rolling cosmologies are themselves dynamically stable is governed by their full fluctuation problem rather than by the vacuum window alone.

These results open several promising avenues and reframe long-standing questions. Because the two members of a pair differ only in the goldstino metric, which in string compactifications encodes the coupling of an anti-D3-brane to the bulk moduli \cite{Kallosh:2014wsa,Bergshoeff:2015jxa,GarciadelMoral:2017vnz,Kallosh:2017wnt}, determining whether and how both $h_+$ and $h_-$ arise microscopically would elevate the correspondence to a direct relation between string vacua. This same structure brings to completion the direction initiated in Ref.~\cite{Atli:2020dni} and isolates the physical content of the de Sitter swampland conjecture within this class: since both members share the visible sector and $|V|$ with all its normalized derivatives, the conjecture, trivially satisfied on the AdS side, constrains only whether the over-uplifted goldstino metric $h_+$ admits an ultraviolet completion, while the bounded regime $0\lesssim-\eta_V\le c'$ of Ref.~\cite{Atli:2020dni} is now realized without ghosts on the healthy AdS partner along the image of the inflationary trajectory. Conversely, since the AdS member always breaks supersymmetry, any ultraviolet completion of a light-class pair directly confronts the conjecture that non-supersymmetric AdS vacua are unstable \cite{Ooguri:2016pdq} against the existence of metastable dS vacua. Finally, the weight identity (\ref{eq:Delta}) and the domain-wall/cosmology map invite a deeper holographic study of the dS member along the lines of Refs.~\cite{Strominger:2001pn,McFadden:2009fg}, extended to full cosmological perturbations, curved slicings, and quantum corrections. Already at the classical level, this correspondence places de Sitter and anti-de Sitter vacua, slow-roll inflation and holographic domain walls, and the uplifted moduli of string compactifications into a single ghost-free supergravity framework, allowing each to be understood in the language of the other.

\section*{Acknowledgments}
The question at the heart of this work goes back to the beginning of the author's doctoral studies in 2017, with an early step taken in collaboration with U\u{g}ur Atl\i~\cite{Atli:2020dni}, whom the author warmly thanks. Although various AI tools in recent years helped test ideas and refine the presentation--assisting with symbolic and numerical cross-checks, code verification, and manuscript editing--the missing piece that unlocked the correspondence was forged through years of human persistence, with the final spark struck, the author is happy to report, in carbon rather than in silicon. The author takes full responsibility for the content of this work.

\appendix
\section{Covariant Potential Reversals and Sign Lock on \texorpdfstring{$W=0$}{W=0} Loci}\label{app:W0}

This appendix shows that on loci where the superpotential vanishes no covariant rearrangement of the K\"ahler potential and the superpotential produces two ghost-free theories with opposite potentials.

Wherever $W=0$ the term $-3e^K|W|^2$ is absent and the potential is given by Eq.~(\ref{eq:W0pos}). Many inflationary models use a stabilizer field $S$ with $W=Sf(z)+\mathcal{O}(S^2)$ \cite{Kallosh:2010xz}, so that $W=0$ along $S=0$. Writing $K=k(z,\bar z)+G(z,\bar z)|S|^2+\mathcal{O}(|S|^4)$, the potential along $S=0$ is
\begin{equation}\label{eq:W0lock}
V\big|_{S=0}=\frac{e^{k}|f|^2}{G},
\end{equation}
and its sign is the sign of the stabilizer metric $G$. If $S$ mixes with the visible fields, $G$ is replaced by the Schur complement $G-K_{S\bar b}\,k^{a\bar b}K_{a\bar S}$ of the visible block, with $K_{S\bar b}\equiv\partial_S\partial_{\bar b}K$ at $S=0$, which is positive whenever the full metric is. The upper block of table~\ref{tab:signs} is the special case of a nilpotent stabilizer, $S=\Xh$, with $G=h=\sigma_2$.

One may try to reverse the potential by the naive rule $(K,W)\to(-K,W)$, which is not covariant under K\"ahler transformations $K\to K+\xi+\bar\xi$, $W\to e^{-\xi}W$. Starting from
\begin{equation}\label{eq:Kminus}
K_-=k-G|S|^2,\qquad W_-=Sf,\qquad V_-=-\frac{e^k|f|^2}{G}<0,
\end{equation}
with $G(z,\bar z)>0$, which has a ghost in the $S$ direction, two K\"ahler-covariant choices reverse the potential exactly. With a new stabilizer $Y$, evaluated along $Y=0$, and constants $C_{\rm dir}$ and $C_{\rm rec}$, the direct and the reciprocal choice are
\begin{align}
K_{\rm dir}&=-k+Ge^{-2k}|Y|^2,\qquad W_{\rm dir}=C_{\rm dir}\,Yf,\qquad V_{\rm dir}=-|C_{\rm dir}|^2\,V_-,\label{eq:direct}\\
K_{\rm rec}&=-k+G^{-1}|Y|^2,\qquad\ \ W_{\rm rec}=C_{\rm rec}\,\frac{Y}{f},\qquad\ \ V_{\rm rec}=-\frac{|C_{\rm rec}|^2}{V_-}.\label{eq:reciprocal}
\end{align}
Both choices are K\"ahler covariant: under $k\to k+\xi+\bar\xi$ and $f\to e^{-\xi}f$, with the source stabilizer held fixed, the new stabilizers transform as $Y\to e^{2\xi}Y$ in the direct and $Y\to Y$ in the reciprocal choice, and both target pairs $(K,W)$ then undergo the K\"ahler transformation $K\to K-\xi-\bar\xi$, $W\to e^{\xi}W$; the reciprocal choice requires $f\neq0$. In both cases, however, the visible metric changes from $k_{a\bar b}$ to $-k_{a\bar b}$.

This is not an accident of the construction. Equation~(\ref{eq:W0pos}) is a sign lock: in a two-derivative $\mathcal{N}=1$ supergravity with positive-definite K\"ahler metric and positive gauge kinetic function, a configuration with $V<0$ requires $W\neq0$. A map $V\to-V$ that keeps $W=0$ on both sides, whether it uses the superconformal signs of section~\ref{sec:signs}, the exchange $K\to-K$ or the covariant maps (\ref{eq:direct})--(\ref{eq:reciprocal}), therefore introduces a wrong-sign kinetic term on one side: on $W=0$ loci the sign of the potential and the sign of a kinetic term travel together. The negative-potential member of a ghost-free pair must therefore have $W\neq0$, where the compensator term $-3m_{3/2}^2$ can compete with the supersymmetry-breaking energy. The nilpotent multiplet of section~\ref{sec:construction} is an economical way to arrange this; the sign lock itself does not depend on it.

\section{Stabilizer Models on \texorpdfstring{$W=0$}{W=0} Loci and a Multivacuum Profile}\label{app:stabilizers}

This appendix collects the scalar spectra of stabilizer models of inflation, which have $W=0$ along the inflationary trajectory and therefore no healthy partner with the opposite potential on that locus (appendix~\ref{app:W0}), and a profile of the inflationary pair of section~\ref{sec:multivacuum} with separate Minkowski and de Sitter minima. As in section~\ref{sec:inflation}, $f$ and $\hat f$ are real holomorphic functions, $V=3H_V^2$ on rolling configurations, the slow-roll parameters are those of Eq.~(\ref{eq:slowroll}), and masses are covariant Hessians in canonical normalization.

\subsection{Stabilizer spectra}

In the shift-symmetric model of Kallosh, Linde and Rube \cite{Kallosh:2010xz},
\begin{equation}\label{eq:M1}
K=|S|^2-\tfrac12(\Phi-\bar\Phi)^2-c_S|S|^4-c_\beta|S|^2(\Phi-\bar\Phi)^2,\qquad W=Sf(\Phi),
\end{equation}
with real constants $c_S$ and $c_\beta$, we write $S=(s_1+is_2)/\sqrt2$ and $\Phi=(\varphi+i\beta)/\sqrt2$. The potential along $s_1=s_2=\beta=0$ is $V=f(\varphi/\sqrt2)^2$, with $\epsilon_V=(f'/f)^2$ and $\eta_V=(f'/f)^2+f''/f$, where a prime denotes $d/d\Phi$, and the masses are
\begin{equation}\label{eq:M1masses}
m_{s_1}^2=m_{s_2}^2=3H_V^2(4c_S+\epsilon_V),\qquad m_\beta^2=3H_V^2(2-4c_\beta+2\epsilon_V-\eta_V),\qquad m_\varphi^2=3H_V^2\eta_V .
\end{equation}
Masses $m_{s_{1,2}}^2,m_\beta^2\gtrsim H_V^2$ require $c_S\gtrsim\frac1{12}$ and $c_\beta\lesssim\frac5{12}$ at leading order in slow roll. The healthy branch of the shift-symmetric model of Ref.~\cite{Atli:2020dni} is of this form, with $K=-\K$ as in table~\ref{tab:signs}.

For $\alpha$-attractors with the stabilizer inside the logarithm and a constant quartic coefficient,
\begin{equation}\label{eq:M2inside}
K=-3\alpha\ln\Big(1-\frac{|\Phi|^2+|S|^2}{3}+\frac{c_S}{3}|S|^4\Big),\qquad W=S\hat f(\Phi)\Big(1-\frac{\Phi^2}{3}\Big)^{(3\alpha-1)/2},
\end{equation}
with $\alpha>0$ and a holomorphic function $\hat f$, the potential along $S=0$ and real $\Phi$ is $V=\hat f^2/\alpha$. The canonical inflaton $\varphi$ is defined by
\begin{equation}\label{eq:canonical}
\Phi=\sqrt3\tanh\frac{\varphi}{\sqrt{6\alpha}},\qquad 1-\frac{\Phi^2}{3}=\cosh^{-2}\frac{\varphi}{\sqrt{6\alpha}} .
\end{equation}
For general $\alpha$, $\hat f$ and $c_S$ the stabilizer masses are exactly
\begin{equation}\label{eq:M2mass}
m_{s_1}^2=m_{s_2}^2=3H_V^2\Big[\epsilon_V-\frac{2}{3\alpha}+\frac{4c_S}{\alpha}\Big(1-\frac{\Phi^2}{3}\Big)\Big],\qquad \epsilon_V=\frac{1}{\alpha}\Big(1-\frac{\Phi^2}{3}\Big)^2\Big(\frac{\hat f'}{\hat f}\Big)^2 .
\end{equation}
On the plateau $1-\Phi^2/3\to0$ and $\epsilon_V\to0$, so the stabilizing term disappears and, for $c_S=0$, $m_{s_{1,2}}^2\to-2H_V^2/\alpha$. For the T-model, $\hat f\propto\Phi$, one has $1-\Phi^2/3\simeq3\alpha/(2N_e)$ at $N_e$ e-folds before the end of inflation, so that a tachyon-free stabilizer requires
\begin{equation}\label{eq:cScrit}
c_S>\frac{1}{6(1-\Phi^2/3)}\simeq\frac{N_e}{9\alpha},
\end{equation}
a quartic coupling that grows with the number of e-folds. The suppression is specific to a constant quartic coefficient. In the superconformal $\alpha$-attractors of Ref.~\cite{Kallosh:2013yoa} the quartic term is $c_S|S|^4/[3(1-|\Phi|^2/3)]$, which amounts to replacing $c_S$ by $c_S/(1-\Phi^2/3)$ in Eq.~(\ref{eq:M2mass}). This gives
\begin{equation}\label{eq:M2KLR}
m_{s_1}^2=m_{s_2}^2=3H_V^2\Big[\epsilon_V+\frac{12c_S-2}{3\alpha}\Big],
\end{equation}
positive for $c_S>\frac16$, as found there. In both versions the second scalar of $\Phi$ has, independently of $c_S$,
\begin{equation}\label{eq:M2beta}
m_\beta^2=3H_V^2\Big[2-\frac{2}{3\alpha}+2\epsilon_V-\eta_V\Big],
\end{equation}
in agreement with Ref.~\cite{Kallosh:2013yoa}; on the plateau it requires $\alpha>\frac13$.

With the stabilizer outside the logarithm \cite{Carrasco:2015pla},
\begin{equation}\label{eq:M2outside}
K=-\frac{3\alpha}{2}\ln\frac{(1-|\Phi|^2/3)^2}{(1-\Phi^2/3)(1-\bar\Phi^2/3)}+|S|^2-c_S|S|^4,\qquad W=S\hat f(\Phi),
\end{equation}
the potential is $V=\hat f^2$; the T-model gives the spectral index $n_s\simeq1-2/N_e$ and the tensor-to-scalar ratio $12\alpha/N_e^2$. The masses are
\begin{equation}\label{eq:M2outmass}
m_{s_1}^2=m_{s_2}^2=3H_V^2(4c_S+\epsilon_V),\qquad m_\beta^2=3H_V^2\big(2+2\epsilon_V-\eta_V\big).
\end{equation}
The stabilizer is protected on the plateau, and the term $-2/(3\alpha)$ of Eq.~(\ref{eq:M2beta}) is absent, so that on the plateau $\beta$ is stabilized for every $\alpha$; in general its stability requires $2+2\epsilon_V-\eta_V>0$. Away from critical points the covariant Hessian (\ref{eq:M2outmass}) differs from the coefficient of $\beta^2$ in a coordinate expansion of the potential by a connection term,
\begin{equation}\label{eq:M2coord}
\frac{\partial^2V}{\partial\beta^2}\bigg|_{\rm canonical}=3H_V^2\Big[2+\epsilon_V-\Big(1-\frac{\Phi^2}{3}\Big)^2\frac{\hat f''}{\alpha\hat f}\Big]=m_\beta^2-\frac{2}{\alpha}H_V^2\Big(1-\frac{\Phi^2}{3}\Big)\Phi\,\frac{\hat f'}{\hat f},
\end{equation}
which vanishes on the plateau.

\subsection{A profile with Minkowski and de Sitter minima}\label{app:multivacuum}

Separate Minkowski and dS endpoints of the inflationary pair of section~\ref{sec:multivacuum} can also be combined in one profile. With $q\equiv\varphi/v$ and constants $C>3\Lambda>0$, the slice profile
\begin{equation}\label{eq:mv_asym}
P(\varphi)=Cq^2(1-q)^2+\Lambda q^2(3-2q),
\end{equation}
extended off the slice as in Eq.~(\ref{eq:mv_profile}), has critical points at $q=0$, where $P=0$, at $q=1$, where $P=\Lambda$, and at $q=\frac12(1+3\Lambda/C)$ between them, where it takes its largest value on the segment, $P_b=(C-\Lambda)(C+3\Lambda)^3/(16C^3)$, with $P_b-\Lambda=(C-3\Lambda)^3(C+\Lambda)/(16C^3)>0$. The first is a Minkowski minimum of the dS member and a Minkowski maximum of the AdS member, the second a dS minimum and an AdS maximum, which obeys the BF bound for $v^2\ge8(C-3\Lambda)/(3\Lambda)$, and the third a dS barrier and a genuine AdS minimum. In the embedding (\ref{eq:mv_model}) both metrics are positive on the segment $0\le q\le1$ and in a tube around it when $3M^2>P_b$.

\bibliography{refs}

\end{document}